\documentclass{article}

\usepackage[english]{babel}
\usepackage{latexsym}
\usepackage[usenames]{color}

\begin{document}

\title{Decision Taking for Selling Thread Startup%
}
\author{%
Jan A. Bergstra\\
{\small  Section Theory of Computer Science,
Informatics Institute,} \\
{\small Faculty of Science,
University of Amsterdam, The Netherlands.}%
\thanks{Author's email address: {\tt j.a.bergstra@uva.nl}. }
\date{}
}

\maketitle

\begin{abstract}
\noindent Decision Taking is discussed in the context of the role it may play for a selling agent in  a search market, in particular for agents involved in the sale of valuable and relatively unique items, such as a dwelling, a second hand car, or a second hand recreational vessel. 

Detailed connections are made between the  architecture of decision making processes and a sample of software technology based concepts including instruction sequences, multi-threading, and thread algebra.

Ample attention is paid to the initialization or startup of  a thread dedicated to achieving a given objective, and to corresponding decision taking. As an application, the selling of an item is taken as an objective to be achieved by running a thread that was designed for that purpose.
\end{abstract}

\section{Introduction}\label{sec:Intro}
In \cite{Bergstra2011a} I have proposed an explanation of decision taking and the way it is embedded in and differs from decision making. In \cite{Bergstra2012a} I have investigated the concept of decision taking as a service.%
\footnote{In \cite{Bergstra2011a} it is found that  decision taking can sometimes be outtasked (see \cite{BDV2011b}), but that it cannot be outsourced, thus contradicting  \cite{Valdman2010}.}  In \cite{Bergstra2012b} the distinction between making a choice, action determination and decision taking is investigated. 

Decision taking admits a deeper understanding when taking a specific context into account in which it is supposed to occur. Focus on such a context amounts to a refinement of the theme of decision taking. Refinements can be introduced in successive phases. 
By means of successive refinement to specific contexts in which decision taking plays a vital role the ideas outlined in \cite{Bergstra2011a,Bergstra2012a,Bergstra2012b} can be made more concrete.   Examples of refinements with suggestions for further refinements  can be found easily, her are some examples:

\begin{description}
\item{\em Selling valuable goods.} This theme may be further refined to different classes of goods and their respective markets, such as for instance:
	\begin{itemize}
	\item Housing market.
	\item Market for agricultural land.
	\end{itemize}
\item{\em Human resource management (HRM).} This theme allows many refinements, each admitting further refinement, e.g.:
	\begin{itemize}
	\item Unit management in SMEs (small to medium size enterprises) where 
	each current or potential employee has an 
	individual identity, role, and context and where only few general rules can be applied. 
	Some further options for refinement:
		\begin{itemize}
		\item HRM for an academic research institute.
		\item HRM for a non-profit organization primarily based on volunteer activity.
		\end{itemize}
	\item Middle management in a large organization where conformance to company wide rules and 
	figures needs to be established.
	\begin{itemize}
		\item HRM aspects of organizational change process for a part of a larger organization.
		\item Recruitment activities for a group of IT oriented SMEs  in a metropolitan region.
		\end{itemize}
	\item Top management in a large organization where HRM principles that apply in many different circumstances
	must be laid down. Further refinement may be needed, e.g.:
	\begin{itemize}
		\item Development of the work force in a professional organization (e.g. a hospital).
		\item HRM policy development for an academic institution.
		\end{itemize}
	\end{itemize}
\item{\em Budget allocation.} Budget allocation and financial planning has many instances.
\begin{itemize}
	\item Middle management aspects: ad hoc development of budget allocation models for 
	specialized organizations, and application to actual problems of financial distribution and accounting.
	\item Planning of financial transfer through generations: making appropriate  testamentary agreements 
	in a specific case.
	\end{itemize}
\end{description}
Below I will consider the first of these cases only:  decision taking involved in sales processes for valuable goods viewed from the side of the seller.  A further limitation is to selling processes where the seller expects financial compensation only, and where the seller is largely  indifferent to the identity and the objectives of the buyer. This specialization may subsequently be refined to specific markets as mentioned above, and to many more markets such as works of art, private possessions of famous persons, classic cars, or luxurious yachts.%
\footnote{The paper has been written with the housing market in mind as a further refinement of the theme at hand to which these considerations should apply at least. An initial plan to write about decision taking for sellers on the housing market proved difficult without first developing the intermediate stage as presented in this paper. An unintended bias to that particular application may have resulted, however.

Topic specific theories of decision making abound. An example of a detailed theory  of decision making in a specific topic can be found in \cite{KimBKN2008}. I have not found an intermediate stage between decision making in general and decision making in the theme of child welfare service delivery: a (general) theory of decision taking for social workers.}

\subsection{Objective of the paper}
The objective of this paper is to find a connection between (i) the terminology of so-called outcome 
oriented decision taking (theory), (ii) a refinement of decision taking to the theme of selling thread control and in particular to selling process initialization, and (iii) the terminology of instruction sequences, 
instruction sequence effectuation, execution architectures, thread production, thread algebra,  and multi-threading. 

The motivation for doing so is (i) to find out if, and if so how  thread algebra and the theory of instruction sequences can be made into a helpful tool for analyzing decision taking in a specific context, and in the specific context of selling valuable and somehow unique goods in particular, and (ii) to make progress towards the development of applications of OODT (see below).

\subsection{Outcome oriented decision taking: OODT}
Together the work in \cite{Bergstra2011a,Bergstra2012a,Bergstra2012b} constitutes a novel approach to a theory of decision taking to the best of my knowledge. Whatever the merits of that approach may be, it is practical to have a name for it. I will speak of the ``outcome oriented (view of) decision taking'' (OODT). In OODT decisions are taken by an agent with the immediate objective to produce a so-called decision outcome, which in most cases is an information module, containing prescriptions regarding future activities and expected or intended states of affairs,  the existence of which is supposed to cause its being effectuated thereby contributing to the realization of the intended effects. The effectuation of a decision outcome, however, is not considered to constitute a part of the decision taking process. Reasoning activities which are needed to produce an informed prediction of the consequences of effectuating a  decision outcome (under design) will be considered a part of decision preparation, rather than of decision taking.

In OODT decisions are actions, first of all leading to decision outcomes. Decisions are relatively scarce events
between many more non-decision events. Non-decision event types include: choice,  preference formation,  preference modification,   plan formation,  and action determination. Action determination takes place for instance in real time setting where an agent must choose between different possible courses of action and immediately perform as chosen. Formulating candidate decision outcomes is among the actions that may precede decision taking. The intermediate role of a decision outcome as specified in more detail in \cite{Bergstra2011a} differentiates decisions from other actions.

\subsection{Presuppositions of OODT}
In \cite{Bergstra2011a}, which I will use as the main source for OODT,  it has been argued that (i) a decision is an act of decision taking, performed by an agent (decision taker), operating in a specified role, having explicit intentions, and equipped with an explicit expectation of how its decision will contribute to a realization of these intentions, (ii) a decision produces a decision outcome, which is a tangible piece of information, (iii) the decision outcome may trigger agents in its scope to put it into effect, thus leading to the consequences of the decision outcome, (iv) decision taking constitutes a final phase of decision making, (v) decision taking  plausibly involves carrying out a protocol, the preparation of which is a task of the decision making process, (vi) determination of the content of the decision outcome and of parts of it belongs to decision making (and in particular to what is termed decision preparation in  \cite{Bergstra2011a}) rather than to decision taking.%
\footnote{An example of a protocol element of a decision taking thread occurs  when selling a home (that is just before transferring economic ownership in a formal session): (i) one needs to check that the property has been properly insured by the buyer, (ii) one needs to check having available all keys, (iii) one needs to have available all current metering data concerning water and various forms of energy, (iv) the property has been brought in the required state, (v) one needs to have passports or other means of identification available for all sellers, or otherwise have transferred their representation to someone else who will attend the session, (vi) a bank account has been provided to the solicitor which can accommodate the sum that will be transferred once the legal ownership has been transferred as well.}

The garbage can model of decision making of \cite{CohenMO1972} fits well in this view of decision taking. In that model decision making produces plans that may constitute candidate solutions to forthcoming problems, while decision taking singles out and activates candidate solutions that are considered proper solutions for actual problems, at appropriate moments.

OODT as outlined in \cite{Bergstra2011a,Bergstra2012a,Bergstra2012b} does not provide an abstract notion of decision of which the particular definitions given in OODT constitute an implementation. Providing such abstract requirements is not an objective of this paper either, but it may be useful to assert that a decision, and its decision outcome must constitute somehow a complete story on how to go about concerning a certain theme. This may require setting a number of interrelated quantitative parameters at the same time in a setting where a survey of a full search space or solution space is missing. 

A decision outcome may be compared to a solution of a problem, which need not be understood as a choice from a space of possible solutions, because only a single candidate solution may have been developed and found to be satisfactory, thereby pre-empting the further search for more candidate solutions. A key difference between decision taking and problem solving is that there is no notion comparable to the notion 
 of a problem that underlies decision taking.
 
\subsection{Quality decisions: preparing for later peer judgement}
Although it is tempting to claim that good decisions must betaken, it is quite difficult to indicate what that may mean.
Quality assessment of decisions and decision taking critically depends on the possibility to find appropriate abstractions that may be used to assess a range of decision taking protocols. At this stage I am not aware of such abstractions.%
\footnote{If significant abstractions from a decision type can be found, that fact may be considered an 
indication that the decision type is insufficiently abstract.}

For decisions in connection with selling valuable  goods I will use the following quality criterion. A decision was good, or adequate, if the corresponding decision outcome was good or adequate. Assessment of the quality of a particular decision outcome takes place in successive phases; at various later stages in a decision taker's existence other agents (peers) may evaluate his past decision outcomes. That evaluation often makes use of  additional information of how things actually went after a decision was taken.%
\footnote{For instance $P$'s decision to buy a specific house may be considered mistaken in hindsight by a $P$'s relatives if $P$ never managed to move to the house and subsequently had to sell it prematurely for that reason.}

Performing decision taking in such a way that from a wide range of perspectives and in a wide range of later episodes relevant other persons or organizations the decision outcome is considered plausible and adequate is the chief objective of employing systematic methods as prescribed by OODT to decision taking. Generating and preserving written records of decision outcomes as well as of comprehensive motivations of decision outcomes and of rationales for the expected effects of decision outcome effectuation (each asked for by OODT) are each conjectured to correlate positively with this ``quality measure''. 

In other words: following the suggestions of OODT in the context of a specific theme contributes to the acquisition of the conjectural ability (in the terminology of \cite{BDV2011a}) to produce decision outcomes which will be judged positively (if assessed at all) by a variety of spectators in a variety of later circumstances and episodes.%
\footnote{In economic psychology it has been established that minimizing anticipated regret influences choices made by human agents (see \cite{ButlerHighhouse2000}). Anticipated regret minimization has been coined as a primary cause of so-called inaction inertia (for instance in \cite{ButlerHighhouse2000}),  view which is challenged in \cite{ZeelenbergNPD2006}. Inaction inertia occurs if a decision taker misses an opportunity to take a profitable decision (for instance to accept a bid for a good $G$ which the decision taker is willing to sell), and then in a later stage is at an increased risk to miss another opportunity to take a similar but less profitable decision (for instance a lower but still significant bid for the same good $G$).}

\subsection{Type theory for decisions}
\label{TTfD}
Further development of OODT requires some form of typing of decisions. Type theory for decisions is probably a challenging theme by itself worth of many years of independent research, but  lacking a well-founded theory of decision types some informal patchwork on decision typing is needed at this stage. 

As an introduction to the topic of decision typing one may consider the notion of a landing. When passengers of an airplane hear the familiar phrase ``cabin crew prepare for landing'' they have now doubt: meant is the next landing performed by the airplane that is currently carrying them as its passengers, which landing is supposed to be a ``normal landing'', and for which the destination airport is known at the moment that the quoted command is being issued. The context disambiguates all possible variety in this use of the concept of a landing. It will refer to (at most) one, and therefore unique, event (act of airplane landing),  which is expected to take place in the near future. Precise timing of the landing may yet be unknown (at the time of issuing said command), and perhaps there may be a last minute choice in favor of the use of  another landing strip, thereby deviating from the original flight plan. The obvious fact that the cabin crew is not supposed to prepare for a landing inside a Soyouz capsule comes about from a context dependent typing regime for landings.

Now in terms of types of landing there is much variation possible: landing at airport $A$ (versus another airport), landing of plane $p$ (versus landing of another airplane of the same type), landing with a plane of type $t$ (versus landing of an airplane of another type), landing performed by crew $c$ (versus a landing performed by another crew), daylight landing (versus landing at night), emergency landing (versus normal landing), successful landing (versus crash landing) non-scheduled landing (versus scheduled landing), landing in bad weather conditions (versus landing in unproblematic weather conditions), and so on. 

Of course intersection types matter most: a successful emergency A380 daylight normal weather landing will be world news wherever and whenever it may occur. 

\subsubsection{Decision types and decision type equivalence}
Just as a landing a decision may have many types at the same time. Which type to use depends on the context. For instance if one intends to decide to buy a second hand car, then the type may be ``my own next decision on buying a second hand car''.  Let $K_{shc}$ serve as a name for that decision type. That type leaves open which brand is chosen and when and where the car will be bought. Given a decision type and the expectation that at most a single decision of that type will be forthcoming, it becomes possible to use a name for that decision, say $d$, and to refer to it in spite of the remaining uncertainties, and in spite of the fact that the event may not take place at all. Thus $d$ refers to a decision of type $K_{shc}$ if any.

What I will need in particular is the concept of different decisions having the same type. Taking the example of buying a second hand car one step further: given the type $K_{shc}$, two different decisions of that type are decision type equivalent. The relevance of decision typing becomes manifest when analyzing the role of decision preparation. In Section \ref{WIA} below the process of decision preparation will be dealt with in detail. Consider the case that an agent $a$ prepares for a single decision (say for particular decision $d$ of type $D$). Such decision preparation may either be performed just in time or well in advance. One may also prepare for $K_{shc}$ decisions in general, not yet aware of which particular decision will be taken (for instance by narrowing down the type to buying a four wheel drive with automatic gear with a production date after 2002 and at most 200.000 km of usage). This form of preparation will be well in advance (that is not just in time) and tactical.

One may also prepare more generally for a range of inequivalent decisions of different types. For instance one may contemplate buying either a four wheel drive, or a recreation bungalow, or a recreational boat. These are three types of decisions, but some preparations (such as realistically finding out how much spare time will be available to make use of these seemingly useful goodies)  may be shared for each of these classes. Such preparations will be classified as strategic well in advance preparations, because of the relevance for different types of decisions.

\subsubsection{Decision outcome type (DOT)}
Decision types may convey information about a variety of aspects such as: decision outcome, decision taker, decision taking protocol, decision preparation method, decision timing (urgency). An urgent decision (of type $K$) is taken (or will need to be taken) briefly after the decision taker  became aware that a decision of type $K$ should be taken. A proactive decision (of type $K$) is taken with determination of  its timing lying in the hands of the decision taker, whereas a reactive decision has its timing mainly determined by outside forces, in some case making it urgent as well.

Decision classification according what is to be achieved by taking the decision and putting its outcome into effect is most informative. Below the type ``$X$ thread startup decision'' will be used extensively, with $X$ and objective that is supposed to be (likely to be, or simply preferred to be) achieved during the active period of the thread.

Classifying decisions in terms of the (expected) decision outcome (DOT for decision outcome type) is useful because that classification indicates what the decision will be about in technical terms. 

\subsubsection{Complementing DOT with DPP and DTP}
Besides the DOT, a decision may also be characterized, at some level of abstraction, by its decision taking protocol (DTP) and by its decision preparation protocol (DPP). Both DTP and DPP may be specified in many different ways ranging from informal plan descriptions to detailed instruction sequences. Descriptions of a triple (DOT, DTP, DPP), each provided at corresponding levels of abstraction may serve as decision types.

\subsection{Process structure and OODT}
In addition to a focus on decision outcomes as an intermediate stage between OODT and the objectives of a decision taker, OODT provides a focus on conceptual structure of decision preparation and decision taking without any bias towards views on how persons  or groups observably behave in decision taking contexts.

This might be compared with formal logic, where the question how human agents perform reasoning is less prominent than the question how different forms of reasoning can be conceptualized in the first place. If I suggest (as I will do below) that a sales process may be viewed as a  multi-thread of multi-threads, with brokers in control of top-level multi-threads, it is no way implied that individuals involved in sales processes are expected to be aware of multi-threading in general or of any formal theory of multi-threading (such as \cite{BergstraMiddelburg2007}) in particular. 

Similarly, if reasoning is analyzed in terms of propositional logic or predicate calculus and a variety of corresponding proof systems it is certainly not implied that rational individuals are supposed to be aware of the concepts predicate calculus syntax, or of any particular syntax for that purpose. Neither is it assumed that individuals showing rational behavior are aware of formal proof systems.

The relation between theory and application in the case of OODT is made more difficult to appreciate because a comparison with modeling in theoretical micro-economics brakes down in spite of a superficial similarity. If the behavior of participants on a market is analyzed mathematically in a simplified model no one hypothesizes that market participants (in the real setting from which the model is intended to constitute an informative abstraction) consciously adhere to and technically understand the model, even if their behavior conforms to its predictions.%
\footnote{It is rather like in physics where the fact (or rather hypothesis) that electrons comply with certain laws is not ``explained'' by requiring that said electrons ``understand these laws''  in a cognitive sense.}
When logic enters the scene, the notion of prediction often becomes an afterthought while conceptual analysis takes priority. While always taking into account that if a dedicated logical theory is likely to be unhelpful for the explanation of human (or group) behavior, it may at the same time be very helpful for the design of artificial agents.

Where logic has a focus on sentential structure and on the structure of reasoning patterns, process theories such as process algebra and thread algebra suggest a focus on process structure. 

\subsection{The role of promises}
In the theory of promises advocated in \cite{Burgess2005,Burgess2007} a promise issued by an agent $u$ informs other agents about what to expect from $u$'s future behavior without engaging $u$ in an obligation. Nevertheless after receiving a promise body issued by $u$ (with $a$ in its scope) it is plausible for $a$ to adapt his expectations of $u$'s behavior accordingly. 

In particular a promise may be stipulated by its promiser  $u$ as being conditional on  a certain event or condition. Reception of  a decision outcome (or a fragment of an decision outcome if outcome fragmentation is applied) by $u$ which was produced when some agent  $a$ took a decision with $u$ in the scope of the broadcast of the decision outcome, is a condition which $u$ may impose on the keeping of a promise he made. The assumption is that $a$ has received the promise body as a promisee from promiser $u$ before taking his decision. Once the decision outcome of $a$'s decision has been received by $u$, is is plausible for $a$ to believe that $u$ will (according to the promise, and with an expectation value related to $u$'s reputation) behave as described in the promise body.

Promises made by other agents constitute an effective mechanism for $a$ when he needs to establish a context where the consequences of putting a decision outcome into effect are sufficiently predictable. Sufficient predictability of the consequences of decision taking, at least in the  decision taker's perception, is in fact a requirement for the concept of a decision.

This assessment of the role of promises indicates a particular class of promises for which the role can be explained in terms of their interaction with decisions. Behavioral promises provide information about the promiser's future expected behavior. Behavioral promises conditional on a promisee's decision, provide such information to a decision taker who is operating in the role of a promisee from the promiser's perspective. Slightly abbreviated it is reasonable to speak of ``decision taking dependent behavioral promises'' in this case.%
\footnote{This phrase expresses two aspects of the matter: (i) the fact that these promises are conditional on the occurrence of specific decisions, and (ii) that the very concept of this kind of promise may depend on the concept of decision taking. Indeed I hold that decision taking can be understood at least in part without reference to the concept of a promise, while in the presence of the concept of a decision it is obvious how to look at this particular class of promises. The notion of a promise as a stand alone concept which serves in that capacity as the starting point of a theory seems to be problematic, however.}

\subsection{Some risks connected with OODT as a research theme}
Having formulated OODT in \cite{Bergstra2011a} and further developed in \cite{Bergstra2012a} and \cite{Bergstra2012b}, 
I am in a position which is both comfortable and uncomfortable. The comfortable aspect is that a range of case studies can be undertaken in order to find out to which extent OODT applies in specific areas and hopefully a contribution to decision taking, both theory and practice,  for those areas can be achieved. 

The less comfortable aspect lies in the presence of various risks, to mention: (i) the risk that OODT cannot be maintained  because its assumptions gradually turn out to be counterproductive or worse, incoherent, (ii) the risk that OODT is merely a rephrasing of an existing view on decision taking
which eventually necessitates a range of modifications in terminology for all subsequent work done on the basis of OODT, (iii) the risk that the distinction between decision taking and decision making, (as expressed in \cite{Bergstra2012b} by means of the equation ``decision making = decision preparation + decision taking'') proves untenable, the risk that the use of terms proves too inconsistent with existing literature. For instance in \cite{Howard1988} decision and outcome are also contrasted but outcome is take to be the result of effectuating the (OODT style) decision outcome.

\section{Decision preparation}
\label{WIA}
According to OODT decision taking constitutes a concluding phase, or activity, of the decision making process. 
This state of affairs may be expressed by the equation ``decision making = decision preparation + decision taking''. Now decision preparation may include many different activities which are performed in different phases of the process. Here is concise classification of preparatory activity according to the distance in time that may be plausible between preparation and application.

\subsection{Proactive decisions and reactive decisions}
Assuming that decisions are classified in types then  a decision taker's ability to determine when a corresponding decision is taken will depend on the type.  Proactive decisions are taken by the decision taker at a time which is under his control, given the type of the decision. Reactive decisions are triggered by external events. Reactive decisions can be urgent because of the need to take them in a timely fashion.%
\footnote{An alternative phrase for proactive decision  open timing decision, and an alternative for reactive decision is closed timing decision.}

\subsection{JIT preparation versus WIA preparation}
It is helpful to classify preparation activities in several kinds related to their urgency relative to a certain kind of decision.
\begin{description}
\item{\em JIT preparation.} Just in time (JIT) preparation takes place immediately before a decision is taken. That is, JIT preparation begins only after the timing of the decision to be taken has been determined. For instance the printing and signing of a decision outcome may be a matter of JIT preparation while the preparation of key fragments of the main text of the decision outcome may have been performed long before.%
\footnote{It is conceivable that such fragments need to be checked by legal and other experts thereby practically excluding JIT mode.}

\item{\em WIA preparation.} Well in advance (WIA) preparation is performed before a corresponding decision is taken and before its precise timing has been determined (thus allowing timing flexibility for the decision taker). WIA preparation is necessary only for reactive decisions, in particular it is needed if, once the need for a certain kind of decision has become clear, the time left is insufficient for a certain kind of preparation.%
\footnote{Repairing a dysfunctional  car stereo system may be done in WIA mode as a preparation for a decision to go on vacation, while repairing a flat tyre can be done in JIT mode, at least by sufficiently  self supporting drivers. Of course WIA preparation my be useful for proactive decisions too, for instance in order to allow the decision taker to react quickly (though freely) on a wide range of external events.}

For a proactive decision all preparations can in principle be performed in JIT mode for the simple reason that the decision taker can afford to wait until all preparations have been made. Given a family of mutually disjoint decision types $K_1, ....,K_n$ a useful distinction between tactical and strategic WIA preparation can be made.
\begin{description}
\item{\em Strategic WIA preparation.} Strategic WIA preparation prepares a decision taker for activities which may be needed for taking a range of possible decisions from different types (say $K_1, ....,K_n$). 
\item{\em Tactical WIA preparation.} Tactical WIA preparation prepares a decision taker in the course of  producing a decision for single type of decision, say a $K_i$ decision.
\end{description}
\end{description}

\subsection{Implied decisions}
\label{ImpliedDec}
Some proactive decision types create, once a corresponding decision has been taken, and its outcome is being effectuated an episode during which (taking decisions that belong to) some decision types may subsequently 
be forced upon the decision taker, in the sense that at various stages during the effectuation of the decision outcome reactive decisions of one or more types must urgently be taken.

Decisions (decision types)  that come unavoidably with the (effectuation of the decision outcome of)  a decision are implied decisions (implied decision types) of that decision. It is plausible that a proactive decision brings with it a range of reactive decisions, as well as proactive decisions (for instance strategy changes if decision outcome effectuation fails to deliver). 

Preparations for implied decision that are incompatible with the degree of urgency of the decision (type) are better viewed as tactical WIA preparations to be included in the preparation of the original proactive decision. 

\subsection{Task oriented thread startup decisions}
Both strategic WIA preparation and tactical WIA preparation are extremes which allow for many intermediate forms. Below I will focus on threads dedicated to achieving specific goals, such as selling a valuable object, or selling a plot of agricultural land. Let $X$ be a type of task and a ``task $X$ thread'' be a thread the  running of which until a particular kind of successful completion, as effectuated by $B$,  is meant to describe (and prescribe in an algorithmic sense) the process of $B$ achieving the objectives implied by task $X$.

When an agent $a$ takes a task $X$ thread startup decision $D_s$, with the intent to have agent $B$  put the instruction sequence defining that thread into effect once it has been started up, the preparation of $d_{tXts}$ of type $D_{tXts}$ may involve tactical WIA preparation for a number of (types of) decisions that $B$ may require $a$ to take during the run, in cases where JIT preparation as a part of the run is not plausible. Agent $a$ may need the acquisition of relevant framework competence as a tactical WIA preparation for the $d_{tXts}$ type decision, whereas the methodology of deriving decision relevant information from the data produced through a diversity of inhomogeneous investigations is likely to be acquired at a strategic WIA level.

\subsection{Plans and planning}
\label{Plans}
In relatively simple cases the sequence of events needed for making a certain kind of decision, say a decision $d$ of type $D$, may easily have the following complexity: 
\begin{enumerate}
\item Strategic WIA preparation (perhaps including the preparation of a script for taking decisions of type $D$ as well as for some other decision types, as well as defining DOTs for decisions of various types including $D$).
\item Tactical WIA preparation for a type $D$ decision by $a$. These preparations  may include the collection of promises (promise bodies) with slow expiration that have been made by other agents conditional upon an impending decision of type $D$  by $a$.
\item Initiating decision preparation for a (proactive) decision of type $D$, including JIT preparation for a decision of type $D$. JIT preparation may include the collection of promises (promise bodies) made by other agents conditional upon an impending decision of type $D$  by $a$. Some promises are more short lived than other promises, which makes their collection more plausible for the JIT stage. Long lived promises may be better collect in tactical WIA mode.
\item (As a part of the JIT preparation for $d$)  tactical WIA preparations for implied decisions  that come along with decision type $D$. 
\item Performing a risk analysis for $d$, involving all known facts.
\item Actually taking a decision $d$ of type $D$ (possibly by effectuating a script that has been prepared for it in either JIT or WIA mode). The decision taking action may require last minute risk assessment (with the secondary risk of preemption of the decision taking activity).%
\footnote{Checks made during decision taking are subsumed under LMA (last minute activity) connected with the decision.}
\end{enumerate}
Now all of this is unlikely to occur without some form of planning in advance done by or for prospective decision taker $a$. This is a subtle matter: a plan $p$ that has been designed (by or for agent $a$) for helping $a$ with making a decision of type $D$ will include the act of taking a $D$ type decision. But that suggests that an agent who is committed to (the effectuation of plan) $p$ somehow implicitly has taken de decision of type $D$ already. This view in turn 
effectively moves the 
type $D$ decision forward in time which is not plausible for the OODT idea of what a decision is namelys a timed action by a known actor.

A solution for this seemingly paradoxical state of affairs is that the plan at hand may at several stages be preempted.
Indeed, although the plan viewed as  an instruction sequence contains the type $D$ decision ``as an instruction'',  that part of the plan need not always be put into effect, in so that the corresponding decision may fail to occur. What seems to be unavoidable when plans (for taking decisions) are introduced is this:  that a situation may arise in which before a certain decision has been actually taken the fact that that will eventually be done has already become unavoidable. Indeed, so-called decisive events during or before decision preparation may create a state in which a decision will necessarily be taken, before it actually has been taken.

Putting a plan into effect may progressively create a situation that makes a certain decision both possible (by having been properly prepared) and unavoidable (after all options of preemption have been missed)

\subsubsection{Terminology for plans and planning}
There is so much correspondence between plan making and decision making that an attempt to develop a terminology for plans compatible in style with the OODT terminology is reasonable. It might be considered necessary given the observation that decision making of moderate complexity cannot be analyzed in detail without having some notion of plan available. Here is a proposal for an OODT style terminology for plans and planning.

A plan equals a planning outcome. A planning outcome results from (an act of) plan taking. Plan taking is an act of planning. A plan can be taken by a single agent or by  a group of agents. Plan making comprises plan preparation plus plan taking. Plan preparation may involve the production of one or more candidate (concept, preliminary, draft, or non final) plans, from which the plan outcome is selected during plan taking. A plan can be effectuated. The plan taker (also called planner, planning agent, or plan taker) is committed to putting the plan into effect once it has been taken. A plan need not be captured by way of a physical representation, it may exist in the mind of the plan taker (or plan takers). This is a difference with decision taking. A planning outcome may be intangible, whereas a decision outcome must be tangible'.

A second difference between decision taking and plan taking is that in plan taking the role of the planner is irrelevant. He simply becomes committed to his own plan.%
\footnote{Of course an agent can decide that a candidate plan that had been prepared before  is to be effectuated. In that case it is clearer to refer to the plan as a task.}

A process that is produced by putting a plan into effect may be called plan guided. Complex decisions are best be taken in the final stage of a plan guided decision making process.

\subsubsection{Decision making as plan effectuation}
It is reasonable to a assume that an agent (or group of agents) takes the plan to make a proactive decision of a particular type, say $D$.
That means that effectuation of the planning outcome will bring about decision preparation as well as decision taking. 

For this way of planning the decision process that is expected to culminate in taking a decision of type $D$ it is essential that the agents involved in plan taking can in principle take decisions of type $D$. In terms of OODT decisions of type $D$ must be in the decision interface of these agents.

A rationale for having a plan the effectuation of which produces a decision making process is that systematic work on plan effectuation can start and only in a later stage it becomes clear whether or not the intended decision is actually taken. 
This setting is reasonable in cases where decision taking is likely to have irreversible consequences, for instance signing a highly consequential contract. 

\subsection{Plans versus threads}
When investigating a particular type of decisions attention must be paid to both the context in which decisions are taken, that is the protocols in which decision play a role,  and the protocols needed, advised, or otherwise conceived to take appropriate decisions.

The decision taker must be able to assume with some grounds that upon a decision being taken, the effectuation of its 
decision outcome brings its objectives nearer to realization. These grounds can be found in the structure of a model of the context as used by the decision taker, together with some assumptions about the validity of that model. One way to formulate such an assumption is to assume that agents involved in the effectuation of the decision outcome will act in predefined ways. That  assumption can be caught in a model which suggests such agents to produce a thread from an instruction sequence that comprises the plan at hand. 

Thus I will use plan as an informal tool for the decision taker to organize the decision making process, whereas I will use thread (as predictably produced from an instruction sequence), to describe how the decision taker may think of the organization of the process of putting the expected decision outcome into effect. Plan has a looser meaning  
than thread in the sense that the plan is only a help for the decision taker and no predictions are based upon it, whereas thread suggests a high (though in practice limited) degree or predictabilty.

Regarding threads, I will make use of some terminology rooted in computing, in particular the thread 
algebra of \cite{BergstraMiddelburg2007} and the theory of instruction sequences (see \cite{BergstraLoots2002a,BergstraMiddelburg2012b}). In particular I will assume that various 
agents (actors, market participants) behave in predictable ways as if they were effectuating an instruction sequence which is known in advance. For description of such behavior I will be using following terminology:
\begin{description}
\item{\em Thread.} A thread is a process (say as in \cite{BaetenBastenReniers2009}) with a reactive structure for its actions (thread actions trigger responses (replies) from the environment in which the thread is being produced). Threads (following \cite{BergstraMiddelburg2007,BergstraMiddelburg2011}) have a binary branching structure because only boolean replies are taken into account. A thread captures uncertainty about the future in its branching structure. For that reason thread are a tool for specifying future behavior. Past behavior fits the model of threads because threads may be deterministic which means that no branching occurs. A thread without branching is also called a trace.
\item{\em Effectuation.} Agent $x$ may put into effect an instruction sequence (alternatively named program, see \cite{BergstraMiddelburg2012}, or script, see \cite{LeighMcGraw1989} and \cite{HummelHNHKL2012} where scripts are used in a multi-threading context, or conceptual model in the terminology of
\cite{ZeeHR2012}), and thereby produce a thread. A thread is the behavior an instruction sequence $p$  that is being effectuated. The agent who enacts the effectuation of $P$ is also referred to as its effectuator.
\item{\em Thread uses service.} A thread is produced in an environment which may contain one or more services (interactive components). Partial evaluation of a thread by composing it with a service produces a further evaluated thread. This composition principle  is formalized by means of the use operator which takes a thread and a service and returns the thread constrained by it having been in interaction with the service.%
\footnote{Complementary to the use operator (which forgets the service after it has been made use of) the apply operator returns the state of the service that has been made use of and forgets the thread.}
\item{\em From thread to trace.}
One ore more successive applications of the use operator can turn a thread into a trace (a deterministic thread), thus formalizing how what might happen in the future turns into what has happened by incorporating an increasing amount of information about the behavior of components (services) with which the thread is interacting.
\item{\em Non-determinism.} Agents have more freedom with the effectuation of instruction sequences than computers commonly have. This aspect may be held against the metaphor of threads as an explanation of agent behavior.
\item{\em Effectuation architecture.} Threads are produced within an effectuation architecture. That architecture prescribes a family of components (called services in thread algebra) the behavior of which a thread may control or influence. An agent who is supposed to take decisions may operate in the role of a service seen from the perspective of a thread that is effectuated by a second agent in another role.%
\footnote{Below I will model the interaction between a selling agent $a$ and his broker $B$ as follows: $B$ produces a thread $t$ from an instruction sequence (thus representing the idea that, as viewed from $a$'s perspective, $B$'s behavior is somehow mechanical and predictable), while $a$ constitutes a ``service'' (component used by $t$ to 
produce (by way of partial evaluation) a more refined thread $t^{\prime}$. The latter thread, which describes the combined activity of $a$ and $B$ is applied to other components that encapsulate (i)  the financial position of $a$, (ii) the activities of various listing services, (iii) the activities of potential buyers, and (iv) operations applied to the object that is being sold.}
\item{\em Single agent multi-threading.} Within a thread produced by an agent different ``sub-threads'' may be concurrently active. Thus an actor may be producing a multi-thread by effectuating an instruction sequence that involves instructions for thread creation.%
\footnote{In the case of a sales process, a broker may view the activation of different marketing options as a multi-thread
of thread each describing the activation of a single marketing mechanism.} 
\item{\em Multi-agent multi-threading.} A single agent may need the concurrent activity of a plurality of other agents, each of whom are producing a thread (or single agent multi-thread). The the agent will be under the (probably self-inflicted) influence of a multi-agent multi-thread.%
\footnote{In the case of selling a good, a seller/owner $a$ may ask different brokers to sell it independently and concurrently, thus giving rise to a multi-agent multi-thread of single agent (multi-)threads.}
\end{description}

\section{Process architecture for selling a valuable good}
It will be assumed that agent $a$, (to which I will refer as ``him'', rather than her or ``it'') owns a valuable good $G$ of entity class $C_e$. It will further be assumed that holding $G$ provides $a$ with some utility $u_{a,G}$, say per unit of time, the consumption of which has been the primary benefit for $a$ that comes with ownership of $G$. Besides experiencing the utility $u_{a,G}$ agent $a$ incurs cost, or more generally a disutility, say $du_{a,G}$ per unit of time, for holding $G$. The disutility includes financial as well as other costs.

Ownership and property are complex notions and indeed the expected holding period for a good of class $C_e$ may be so long that property
rights change in between, so that long term valuation by means of discounted expected utility becomes problematic (e.g. see \cite{Cribbet1986}).

Now $a$ may for some reason come to believe that utility and disutility are out of balance  or will be out of balance in a foreseeable future. If $u_{a,G} < du_{a,G}$ then contemplating the termination of the ownership status of $a$ is plausible. Selling $G$ is a plausible method for terminating $a$'s ownership of $G$. Other strategies to improve the situation may involve increasing $u_{a,G}$, or decreasing $du_{a,G}$, or a combination of both.

Selling $G$ by $a$ involves (i) finding another agent $b$ who is willing to exchange ownership of $G$ for an 
amount $P_t$ at time $t$ and under conditions $c_t$ in such a way that ``the deal'' satisfies $a$'s preferences, and (ii) successfully performing the transaction (exchange of $G$ for $P_t$ between $a$ and $b$ at time $t$) as specified.

Although the analysis made in this paper may seem rather abstract I have used the example of an owner occupier selling his home in a difficult market, that is a buyer's market, as a leading intuition. Obviously when selling second hand kitchen equipment  in a local 
amateur market for a good purpose the corresponding decision taking will be  much simpler, if occurring at all. And a seller of pure gold will deal with other uncertainties making some of these considerations irrelevant in that case.

\subsection{A selling thread and its startup decision}
The process structure that I will assume for the sales process of a valuable good $G$ by its owner and prospective seller $a$ is that of a so-called selling thread which is produced by a broker who puts an underlying  instruction sequence into effect. 

For the same good a plurality of selling threads may be simultaneously active, each being produced by different brokers. The selling process for $G$ for that reason may consist of a multi-thread. 

\subsubsection{The role of a broker}
The idea is that a broker, say $B$, acts as an operator in charge of realization of the selling thread. Thus the broker is in charge of thread control. At important phases the  broker needs to involve the owner/seller $a$ by asking him for a decision, the outcome of which is likely to have decisive influence on thread control. Many actions, however, can be independently planned and performed by the broker. It is plausible to assume that the thread is produced by putting an instruction sequence into effect. That instruction sequence contains steering fragments according to the terminology of \cite{Bergstra2010}.%
\footnote{In the terminology of thread algebra \cite{BergstraMiddelburg2007}) and viewed from the perspective of the broker (or rather the thread operated by the broker) the owner/seller $a$ is a ``service'' who he (the broker) may call for decisions. These calls appear in steering fragments (see \cite{Bergstra2010}) in the instruction sequence that underlies the thread.

The comparison with \cite{Bergstra2010} is somewhat defective because in that paper decisions have boolean outcomes and in OODT much more information is needed for a decision, but as far as impact on thread control is meant the comparison is valid and informative.

The boolean reply that goes with a decision is likely to have immediate impact on thread control, all other information which goes with it according to OODT (though ignored in \cite{Bergstra2010}) is likely to impact the way in which the responsible broker operates the selling thread subsequently when it comes to choices and action determination for matters of secondary importance.}
An instruction sequence from which the selling thread is produced will implement an algorithm that depends on the conventions of the broker. In some cases a broker will search for unknown clients and concurrently, in a separate subthread, directly approach in a sequential fashion known clients who have in a prior stage informed the broker of their interest in buying one or more objects of type $C_e$.

A broker $B$ operating on behalf of $a$, who may in fact  be identical to $a$ in a ``simplified'' case, is made responsible  for the effectuation of an instruction sequence (program) thus producing the selling thread. Decision taking by $a$ influences the control of the selling thread. Most importantly $a$ may take a decision to startup a selling thread, but thread termination and intermediate modifications of parameter settings also require decisions taken by $a$. The selling thread may in turn consist of a multi-thread with different marketing options (including so-called listings with listing services) each under the control their own thread.

\subsubsection{The basic action interface of a selling thread depends on $C_e$}
The actions performed while running a thread constitute its so-called basic action interface. That interface may differ depending on the entity class $C_e$ of $G$. If $G$ is a home, the broker may provide information, show potential buyers around, provide them information about local conditions and regulations, solicit proposals from mortgage lenders to demonstrate the feasibility of deals to potential buyers and play a role in the exchange of bids and reactions to bids. If a bid is accepted and no escape from bid acceptance occurs, the final phase of the thread involves execution of the transaction, a phase in which thread control may in some cases have been transferred to other agents, for instance to a notary or to bank representatives.

\subsubsection{Selling thread startup decision for $G$}
A run of a selling thread  for a good $G$ consists of a systematic and planned sequence of actions 
intended to make a sale  of $G$ happen under conditions which serve as parameters of the selling thread.

The owner $a$ of $G$ may take the decision to start a selling thread for $G$. I will call such a decision a
 selling thread startup decision. The decision outcome type (DOT)  of a selling thread startup decision will be discussed in detail in \ref{DOTsts} below.  Requirements on the DOT for selling thread startup decisions can be determined more firmly if the property class $C_e$ can taken into account.
 
Once $a$ has taken a selling thread startup decision for $G$ it is plausible to refer to $a$ as the prospective seller (of $G$). A prospective seller may turn into an (actual) seller by performing a selling transaction.

It is plausible that $a$ has different selling threads for $G$ running concurrently, thus bringing multi-threading into play
(see e.g. \cite{BergstraMiddelburg2007}).

It is probably common to speak of a sales process rather than of a selling thread. By speaking of a thread I intend to highlight the mainly sequential and algorithmic character of the actions involved. By using ``selling'' instead of ``sales'' I intend to highlight the real time character of the activity, rather than to make a reference to an abstract activity class. A selling thread is supposed to consist of the putting into effect (see \cite{Bergstra2011}) of an instruction sequence (see \cite{BergstraLoots2002a,BergstraMiddelburg2012}) in an appropriate execution environment (see \cite{BergstraPonse2007a}).%
\footnote{In the absence of guidance from a sequential algorithm it is preferable to speak of a sales process. Modeling such processes may for instances be done by means of process algebra (see \cite{BaetenBastenReniers2009}).}
It will be assumed that a selling thread  is put into effect by a broker. If no broker is present the broker will be a second role played by the owner $a$.

\subsection{Plan guided decision making for selling thread startup}
A selling process for valuable object $G$ consists of an open ended spectrum of activities by various agents, each of which are somehow connected to a successful or unsuccessful attempt by its owner $a$ to sell the good.

It is assumed that $a$ takes the plan to try to sell $G$, via the startup of a selling thread. Then $a$ effectuates the resulting plan outcome thus producing a decision making process that may culminate in the intended selling thread startup decision (below abbreviated to STS decision).
The following stages will be distinguished:
\begin{description}
\item{\em Strategic WIA preparations.} 
	\begin{description}
	\item{\em Acquiring specialized market competence.}  Developing specialized market 
	competence concerning the 
	relevant market.\footnote{This task may include, $a$ becoming acquainted with the 
	theory and application of subjective probability and the use of Bayesian statistics 
	and Bayesian networks if the impact of recent transactions of an expected selling 
	price must be determined, given less recent but more official estimates that may 
	serve to determine prior odds.}
	\item{\em ASIL development.} Developing an application specific informal logic (ASIL)
	 in order to derive rules of 
	plausible behavior from data about the relevant market in the past and about 
	similar markets in previous periods and perhaps elsewhere (if location matters).
	\end{description}
	
\item{\em Tactical WIA preparations.}
	\begin{description}
	\item{\em DOT definition.} Specifying the DOT (decision outcome type) for the selling thread startup decision.
	\item{\em $C_e$ market framework competence.} Developing framework competence about the $C_e$ 
	market to the extent that either choosing a broker and subsequently cooperating with the chosen 
	broker or playing the broker role is enabled for $a$.
	\end{description}
\item{\em JIT (selling thread startup) decision preparation.}

In general there is no prescription possible for how  the preparation phase needs to be structured. After further refinement to specific markets developing such prescriptions may become feasible.  Decision preparation may contain one or more of the following ingredients.
\begin{enumerate}

\item Choosing an external (that is different from $a$) broker, or choosing not to make use of an external broker.%
\footnote{The activity of a broker may be understood as the outtasking of sales and marketing activity by the prospective
seller. In \cite{Rapp2009} the comparable mechanism of outsourcing of the sales process is discussed. It is indicated that a tendency towards outsourcing sales activity is plausible in the presence of a focus on selling and production as contrasted with a customer orientation or a focus on learning.}
\item Determination of a number of quantities, with reservation price formulation as the most well-known example. This phase may be performed in cooperation with an agent who is expected to act as a broker.

\item Preparing a script (instruction sequence) for taking a selling thread startup decision. 
(JIT preparation for STS decision.)
\item 
\label{Script}
Preparing a script (instruction sequence) which can be effectuated by the broker to produce the selling thread (if and once that has been started). If an external broker is engaged then the script is likely to be a customized version of scripts that he is used to employ for other customers. (JIT preparation for STS decision.)
\item Performing WIA preparations for reactive decisions that are expected to be requested when the selling thread is active. This may involve setting quantitative parameters and writing instruction sequences for the decision protocol.
(Tactical WIA preparation for implied decisions of the STS decision type.)%
\footnote{See \ref{ImpliedDec} for the definition of  implied decisions. Only once the script referred to in item \ref{Script}
is known information is available regarding which reactive decisions are likely to be encountered when the selling thread is running.}
\item Risk analysis, hypothetically assuming that  the STS decision has been taken. (JIT preparation for STS decision.)
\end{enumerate}
During the decision preparation, viewed as effectuation of a plan, that effectuation may terminate in many ways prematurely thus preventing the STS decision to be taken.
\item{\em Selling thread startup decision.} This is a selling thread related decision (taken by $a$). The decision requires checking various constraints concerning values and conditions that have been developed during the decision preparation phase.%
\footnote{These activities are referred to as LMA (Last minute activity) activities fro the decision. A plausible side-effect of LMA is that the decision taking process (or thread if it has been prepared by way of writing an instruction sequence for it) is preempted and as a consequence no decision is taken.}
If the matter is complicated implementing this action of decision taking putting an instruction sequence into effect that has been prepared before is plausible. 

At this stage multi-threading may be in order: $a$ may concurrently take different selling thread startup decisions and have selling threads active, each being produced by different brokers.
\item{\em Selling thread startup.} Caused by the decision outcome of the selling thread startup decision having come into existence the selling thread is initiated by the broker (who may  equals $a$ unless an external broker has been engaged).
\item{\em Selling thread active.} During the active phase of the selling thread $a$ may be asked to take further so-called selling thread related decisions. Some of these are reactive decisions (for instance the bid acceptance decision or bid rejection decision, and some are proactive, for instance the list price modification decision.)
\item{\em Sales transaction effectuation.} Actually performing the sales transaction may or may not be included as a constituent of the selling thread. If not, performing the transaction will appear as a separate stage.
\end{description}

\section{Selling thread related decisions}
The idea that $a$ taking a selling decision concerning $G$ is equivalent to $a$ putting $G$ on the market for some list price that $a$ would be happy to receive in exchange for $G$ any time soon is problematic. Much more is needed to qualify a decision as a  decision to sell, rather than as a mere decision to put $G$ on the market for a phantasy price. Substantial information concerning pricing and timing is required. I will avoid the phrase ``selling decision'' and speak of selling thread related decisions only. 

With selling thread related decisions (or seller side sales related decisions)  I will refer to a class of decisions related to sales processes that can be taken by a selling agent and which affect the run (production) of a selling thread. I will often speak of a prospective selling agent in order to highlight the presence of  uncertainty concerning the fate of the sale.
Below is a survey of selling thread related decisions (or rather decision types) within a proposed naming scheme.

\subsection{Reactive implied decisions}
All selling thread related decisions are implied decisions of the selling thread startup decision. Implied decisions are  either proactive or reactive. Some of the reactive decisions are indispensable. I will provide a brief survey of reactive 
implied decisions (of a selling thread startup decision)  first.

\begin{description}
\item{\em Bid acceptance decision for $G$.}
The decision by owner $a$ to sell $G$ can take place when a bid (equivalently) an offer) has been made by a potential buyer. The decision to sell is identical to the decision to accept the terms of the bid.%
\footnote{Given this meaning of the phrase ``decision to sell'', the phrase ``last week I decided to sell my home and therefore I have contacted a real estate broker today'' is problematic because it is usually assumed that the broker will be helpful for finding a party who produces a bid. The selling decision can only be taken, however, after a bid has been made.} 
The decision to sell will be referred to as the bid acceptance decision.

The difficulty with taking a bid acceptance decision may be that a number of different checks need to be made and that, at least if the the bid is valid for a limited period only, a protocol for doing so needs to be available beforehand. The  selection or development of that protocol is subsumed in the decision preparation stage. It is of the WIA type in the terminology of \ref{WIA}. Several of the quantitative parameters for the protocol must have been estimated beforehand as a part of decision preparation.%
\footnote{An interesting complication arises if due to time delays between decision preparation and decision taking some estimates are outdated. It even may be the case that effectuation of the protocol for taking the selling thread related decision involves putting an instruction sequence (representing some algorithm) into effect which itself contains complex conditions, where during the evaluation of a condition the same quantity is used twice, which so much actions taking place in between that the quantity has changed its value in the mean time. This phenomenon, if ever observed, will bring the proposition algebra of \cite{BergstraPonse2011a} into play in the analysis of selling thread related decision protocols.}

There are several complications with the notion of a bid acceptance  decision:

\begin{description}
\item{\em The sales contract as a joint decision outcome.} In some markets a
 contract that specifies a sale of $G$ between seller $a$ and buyer $b$ may 
constitute the most plausible (or even the only legally valid) form of the bid acceptance decision for the seller at the same time. A written agreement between two agents may be considered a (joint) decision in some cases. Joint 
decisions have not been analyzed in \cite{Bergstra2011a,Bergstra2012a,Bergstra2012b} but the development of a convincing concept of joint decision taking, fit for producing joint decision outcomes, seems to be rather straightforward.
Some further remarks on joint decision are made in Section  \ref{JDec} below.
\item{\em Sales transaction implementation.} The bid acceptance decision may precede the legally sound and complete sales transaction, which by itself may involve a complex thread resulting from the putting into a effect of a specialized recipe (algorithm, instruction sequence) by staff specialized for the implementation of sales transactions for 
entities of class $C_e$.

\item{\em Positive and negative bid acceptance decisions.}
A bid acceptance decision can be positive (the default) or negative if an offer is refused. A negative bid acceptance decision may also be  called a bid rejection decision.

\item{\em Accepting or rejecting a bid without decision taking.} If an offer is made, simply by not reacting within a certain period in many cases the offer can be refused. Also by simply communicating ``yes'' or ``no'' to the potential buyer $B$, or 
to the agent who is representing $b$ if such an agent is present, an offer may be accepted or rejected without that step being the consequence of effectuating (the outcome of) a preceding sales transaction decision. 

\item{\em Selling  without  taking a bid acceptance decision.}
Depending on the conventions in a specific market it may be the case that only a written contract signed by both buyer and seller can be considered the outcome of a (seller side) bid acceptance decision. It follows from the terminology which I am proposing that, say if verbally communicating a mere ``yes'' or ``no'', either directly or via one or more intermediate agents, suffices to accept an offer, that in principle $G$ can be sold by $a$ to $b$ without a (seller side) sales transaction decision to that extent having been taken by $a$. Choosing between expressing yes and no, either in words or with non-verbal physical means, is a matter of action determination rather than of decision taking (see \cite{Bergstra2012b} for that distinction).

\end{description}

\item{\em Bid acceptance escape decision for $G$.}
In some cases a seller has the option to avoid selling $G$ even after a bid acceptance decision for $G$ has been taken. This may occur if some agreed conditions have not been met in an agreed period after the bid acceptance decision had been taken.

\item{\em Call option proposal decision for $G$.}
Rather than to agree with an offer a prospective seller may hope for a better offer from an other potential buyer of $G$. Accepting an offer by way of a corresponding sales transaction decision may puts a premature end to such hopes. 

Rejecting the offer may turn out to have been unfortunate if the hope for a better deal was an illusion after all. If the conventions on the 
specific market for goods of class  $C_e$ allows for contingent pricing (see \cite{BiyalogorskyGerstner2004}) it may be possible for $a$ to respond to an offer (say at price $p$) by offering the prospective buyer $b$ a call option for $G$ against an execution price
equal to $p$ and at a premium $q$ (e.g. 2.5\% of $p$)  to be transferred upon acceptance by $b$ of the deal. The call option expires at some later date which is sufficiently close to keep $b$ interested and sufficiently distant that an
absence of offerings by competing buyers above $p+q$ before the expiration date 
can be understood by $a$ as a sign that $p$ was a good price after all.

It is reasonable to assume that $a$ needs to take a decision (which may be referred to as a call option proposal decision for $G$) before engaging in selling $b$ a call option for $G$ rather than $G$ itself.

\end{description}

\subsection{Proactive implied decisions.}
Proactive decisions are primarily taken if the prospective seller has become dissatisfied with the progress of the selling thread. Proactive decisions are not immediate replies on the behavior of agents different from the prospective seller.
\begin{description}
\item{\em Selling thread termination decision for $G$.}
Part of a selling thread startup decision is information on how to terminate the selling thread. A most plausible reason for a selling thread for $G$ to be terminated is that another selling thread for $G$ has successfully lead to a stage in which the 
prospective seller can indeed take a bid acceptance decision.

It is reasonable, but not necessary to assume that termination of a av thread is a consequence of the effectuation of the outcome of a selling thread termination decision. 

\item{\em Selling thread repositioning decision for $G$.}
While a selling thread is up and running decision preparation for forthcoming decisions may take place. Besides
a selling thread startup decision and a selling thread termination decision, a selling thread repositioning decision may be taken. Repositioning involves a modification of key parameters, with changing the seller reservation price (see below) as a primary example.

\item{\em Run-time broker disengagement decision concerning $G$.} The prospective seller may wish to terminate the involvement of his broker, while the selling thread is being effectuated, perhaps in preparation of transfer of effectuation of the thread (a matter of thread mobility) to another broker.

\item{\em Run-time broker engagement decision concerning $G$.} The prospective seller may wish to engage a new broker. Once a candidate broker has been found who is willing to serve $a$ in that role, $a$ can take a 
decision to that end.

\item{\em Marketing thread startup decision for $G$.}
If a decision is made to put $G$ on the market while $a$ is aware of the (perhaps intentional) fact that it is quite unlikely that $G$ will be sold in compliance with the conditions of a purported selling thread startup decision outcome, the the decision is not a selling thread decision but rather a pseudo selling thread startup decision.%
\footnote{A pseudo selling thread startup decision may be used by $a$ to put $a$'s home $G$ (assuming of course  that $G$ is a dwelling) on the market at a very high price only to help neighbors of $a$ who try to sell at a more realistic price. 

Similarly, but with less friendly objectives, again assuming that $G$ is $a$'s home, $a$ may introduce a problem in $G$ (for instance by renting out a number of rooms to a friend, say $f$), and subsequently put $G$ on the market at a low price (not mentioning that fact that the use o $G$ will be frustrated by the long period of rent agreed by $a$ and $f$ which will be binding for a subsequent owner of $G$ as well. Now $a$ may take a pseudo selling thread startup decision to  put $G$ on the market for a quite low list price while not mentioning the fact that a substantial part of $G$ cannot be used by any new owner in the near future. This in spite of the low list price, no sale is to be expected. An objective of this form of pseudo STS decision  can be to frustrate some of $a$'s neighbors who have recently put similar homes on the market for more elevated prices. This prospective sellers have their offerings seemingly ridiculed by the behavior of $a$.}

A selling thread may consist of a multi-thread involving different marketing threads. This brings hierarchical multi-threading 
and user controlled scheduling of threads into play (see e.g. \cite{BergstraMiddelburg2011}).

\item{\em Marketing thread termination decision for $G$.}
Clearly a marketing thread my be terminated. That may be the case if its cost are considered too high, or if a bid acceptance decision has been taken. Terminating a marketing thread may involve a mere action by the prospective selling agent or bay the broker working for the seller, but it may also require a preceding marketing thread termination decision
the outcome of which is effectuated with said termination of the marketing thread as one of its consequence.

\item{\em Marketing thread repositioning decision for $G$.}
Repositioning of a marketing thread takes place for instance if a list price is modified, or if object feature descriptions are adapted.  It is plausible that marketing thread repositioning occurs as a consequence of effectuating (de decision outcome of) a sales thread repositioning decision. In that case there is no need for an explicit marketing thread repositioning decision. In some cases, however, for instance if the marketing thread startup has not been performed in the course of effectuating (the decision outcome of) a sales thread startup decision, repositioning a marketing thread may need preparatory and dedicated decision taking.

\end{description}

\section{Selling thread startup decision preparation}
In Section \ref{DOTsts} I will outline in detail how a selling thread startup decision outcome might be structured. It will need to carry with it a significant number of quantitative data. These data must first be prepared in the decision preparation phase for the selling thread startup decision.

It is essential that a selling thread startup decision embodies the intention to lead to a sale before the expiration of the reservation period. All aspects mentioned below provide input to the reasoning that $a$ needs to apply in order to understand that it is reasonable to expect that a selling thread startup decision, as well as various subsequent implied decisions will serve the relevant objectives. The logical background of such reasoning processes may be indicated as `` informal logic'', (see \cite{Johnson2006}) Much remains to be investigated about the application of informal logic in specific circumstances, for instance it seems to be an important challenge  to develop an application specific informal logic for selling agents on the housing market.

\subsection{Preliminaries on markets, values and prices}
\label{Constr}
Writing about selling thread startup in full generality for all classes of goods for sale is problematic. Some assumptions and conventions must be taken into account concerning the goods and their market, and about how to speak of values and prices.

\subsubsection{Assumptions on goods and market}
The following assumptions concerning the class $C_e$ of entities in which $G$ resides, and about the conventions that are observed for the $C_e$ market, support the rationale of the preparations sketched  thus unavoidably limiting the generality of this section.
\begin{enumerate}
\item Goods in class $C_e$ have some kind of usefulness or functionality which is kept unaffected when looked after in a normal fashion.

\item Good $G$ of which $a$ is contemplating putting on the market
 is  preserved in a normal and sellable condition by $a$ during the period that it is on the market.
 
\item $G$ is an individual item in its class, its sale is done per item and not in a batch together with other entities of its class.

\item The default assumption is that ownership of $G$ represents some form of wealth, though there must be an open mind for circumstances in which that appreciation is unfortunately invalid.

\item $G$ will be offered to the market by way of an information package about it, probably highlighting its virtues somewhat better than its defects,  in combination with a list price that serves as an asking price. 

\item Offers are supposed to be made by potential buyers who have invested time and energy to getting informed about $G$ before making the offer. 

\item Offers are expected at or below the list price in normal conditions, while in a heated market offers above the list price may have to be handled as well.

\item Offers as well as decisions by the seller to accept an offer are binding, and considerable freedom exists for seller and buyer to customize the details of an offer.

\item Any broker $B$ who is hired by $a$ will be chosen (and indeed can be chosen)  in such a way that he operates consistently with common  conventions of the $C_e$ market. This fact need not be checked by $a$, because it is adequately taken care for by the professional organizations to which $B$ is subscribing.

\item Prospective seller $a$ is free to choose a list price for $G$.
\end{enumerate}

\subsubsection{Value versus price: prices}
A vast literature explains how different values and prices for class of goods may be understood. I will follow a systematic convention on the use of value versus price, knowing that many authors use these terms in different ways.
Value can be used in a financial sense and in a more general sense. I will use value as expressed in financial terms. So both prices and values are expressed in money, say in Euros, together with a date of valuation or pricing. 

A price is always known by way of choice (for instance a list price), measurement (typically a selling price or buying price) or calculation (for instance a cost price, based on a previous buying price, some model of cost of ownership since acquisition and some model for discounting missed income and other opportunity cost). Uncertainties in prices stem from missing data (for instance transaction cost made when buying $G$ many years ago), incomplete or unconvincing models (maintenance cost may need to be separated from genuine improvements, improvements must in some way be separated from adaptations made without the objective  of a lasting contribution to user independent ``property value'').

\subsubsection{Values}
Values, such as market value (what price may be expected when sold), replacement value (what  price will need to be paid,  to replace the good when that needs to be done, corrected for the ``value'' which comes along with having new goods instead of the old ones)  business value (some capitalized form of the contribution to a future income stream, 
and to a corresponding future profit stream that a good provides) are essential figures needed for the control of processes involving a good (assessment of wealth, taxation, inheritance, exchange for other goods, determination of business quality, quality assessment of the result of selling a good, quality assessment for the running of a selling thread and the design of an underlying instruction sequence, quality assessment of a selling thread startup decision including its preparations). 

Values result from expert judgements (about expectation values of prices, or of discounted future income streams)  and will always have two degrees of uncertainty: different experts produce different estimates, and each estimate by necessity has its own variation. In specific markets both forms of variation can be studied empirically.

Thus with the market value (often called fair market value) of $G$  I will denote the expected outcome of  a (possibly hypothetical) sample of experts who each must determine their expectation value of the financial result of selling $G$ in a normal way to a normal buyer under normal conditions, after subtraction of the cost of selling. This is the objective market value of $G$.

The subjective market value of $G$ from the perspective of its owner $a$ is the expected outcome of  a (possibly hypothetical) sample of experts who each must determine their expectation value of the financial result of selling $G$ under constraints specific to $a$ (such as a strong preference of $a$ for a shorter than normal time on market (TOM), or an unusual permissiveness  for an extended TOM, the wish to receive exactly the list price and not to negotiate under any circumstance, a preference to avoid a certain group of buyers, or a preference to sell to some selected potential buyers),  to a normal buyer under normal conditions, after subtraction of the cost of selling. This is the objective market value of $G$.

The subjective market value assumes that a ``normal'' attempt is made  to sell $G$ at the best price under the given selling process conditions as stipulated by $a$. It does not take into account $a$'s needs or wishes in terms of the selling price. In particular if $a$ can easily cope with a yield which is significantly lower than the subjective market value that fact has no downward effect on the subjective market value.

\subsection{Strategic WIA preparation}
It is difficult to find clear distinctions between tactical and strategic WIA preparations independent of the markets on which to sell. Nevertheless in the following paragraphs some options are listed for strategic  WIA type preparation for a selling thread startup decision.

\subsubsection{Choice of competence level}
A prospective seller of $G$ must develop some awareness of possible competence levels regarding the $C_e$ market and then choose the competence level that fits his own purposes best. Even under the constraints listed in \ref{Constr} above, a wide range of options and arguments concerning this matter lies open.
The role of expert knowledge of sellers and buyers concerning a market  seems to be different for different price levels. 
No model of the use of expert knowledge by market participants is likely to cover the whole range of markets. 
The following observations may be helpful for obtaining a perspective on this matter:
\begin{description}
\item{\em Sellers on cheap markets need considerable competence.} Consider the market for second hand 
postal cards, which has many submarkets world wide. On that market sellers and buyers have low margins per sale and sales are 
each unique to some extent. As an implication hiring a professional consultant or broker is less plausible and market participants are likely to be among the best informed individuals available concerning the relevant market.
\item{\em Sellers on markets of moderately expensive items are likely to be less informed.} As an example one may consider that price range of the housing market which contains 80\% of the dwellings in a submarket, that is it omits the lowest 10\% and the highest 10\% of the market. With $C_{hm}^{80\%}(R)$ I denote the 80\% subrange of the housing market in region $R$.

It is uncommon to expect from an owner $a$ contemplating the sale of a home $G \in C_{hm}^{80\%}(R)$ an awareness of the international literature on housing market research. The investment needed for obtaining that awareness would be
considered too high given its infrequent use. Often such market participants are being portrayed as being
amateurs in that same research literature, in spite of the fact that they often make use of brokers (real estate agents, realtors) who must comply with professional standards, and who for that reason cannot be 
considered amateurs.%
\footnote{Medical treatment might also be considered as a market of items of 
moderate cost and buyers on that market are
not expected to develop expert knowledge concerning any  problem unless the problem  occurs highly infrequently, in which cases professional doctors are often not expected to acquire appropriate expert knowledge. The very notion of an amateur seller (service provider) on the market of medical treatments is considered problematic. One might wonder to what extent this very judgement contributes to the exploding costs of health care in the western world. Information asymmetry between client (buyer) an service provider (seller)  is a characteristic feature of the market for medical service. In the housing market that asymmetry is much less present because participants often have selling experience as well as  buying experience.}

\item{\em Brokers active for markets of moderately expensive items are likely to be less informed.} 
Taking the housing market once more as an example. It is probably unreasonable to expect of a real estate agent a systematic awareness of the international literature on housing market research.%
\footnote{It may even be part of the framework competence required of prospective home seller who performs decision preparation for a sales thread startup decision to know that a real estate agent active in the submarket  $C_{hm}^{80\%}(R)$ is likely to be unaware of the available research literature. In contrast a medical doctor working in a specialized hospital will be expected to consult recent specialized literature. The same holds for management consultants in the high-end of the consulting industry.
}

\item{\em Sellers on high end markets markets may be well-informed.} Sellers of  very valuable or complex real estate portfolios serving investment purposes are likely to be aware of recent findings of research concerning their relevant markets. The same holds for sellers preparing transactions at the top end of works of art, top quality music instruments,  exclusive yaghts,  or entire businesses.
\end{description}
The choice of an adequate competence level for $a$ is a strategic matter because it may be the case that so much expertise will be needed that the cost of its acquisition is best amortized over a range of different transactions, that is over its use in a range of non-equivalent decisions. In that case such preparations must be done in strategic WIA mode, which then implies that the preceding choice of a desired competence level is to be done in WIA mode as well.

It is also possible that $a$ determines that the desired level of expertise can be acquired in tactical WIA or in JIT mode.

\subsubsection{Preparing an application specific informal logic (ASIL) for selling $C_e$s.}
It may be the case that the choice of competence level leads the prospective seller to the wish to reason about his proceedings in a way informed by the available research literature concerning the market for entities of class $C_e$.
This objective requires meticulous preparation and it amounts to the development of an application specific
informal logic (ASIL) for (sellers on) the $C_e$ market. The development of an ASIL from scratch, or its acquisition if it already exists, must be done WIA because it may be time consuming and may even have an open ended timing. Because the activity may be quite costly it may be reasonable only in strategic WIA mode in the expectation of a variety of applications for taking different and non-equivalent decisions.

Research papers on particular markets that exist through the world usually provide information on limited and in some sense coherent  submarkets during well-defined  past periods. That information stems from
empirical data and is abstracted into hypotheses which are confirmed by analyzing the degree to which a fit can be found to so-called models, often proposed in the same study, and mostly found by refinement or adaptation of models used in previous studies.   

Each confirmed hypothesis may be turned into a reasoning pattern together with a qualitative or preferably a quantitative estimate of its validity.%
\footnote{Some examples of reasoning patterns that may be obtained this way: (i) a very quick sell of a house is not an indication of selling to cheap (seller mistake), see \cite{SirmansTD1995},  (ii)  in a booming housing market the gap between LP and OESP tends to be higher than in a bust market, see \cite{HaurinHNS2010},  (iii) loss aversion affects not only list prices and expected selling prices but also has an upward effect on transaction prices, 
see \cite{GenesoveMayer2001}, (iv) increasing LP leads to increased TOM, see \cite{AnglinRS2003}, (v)
the number of houses visited by prospective buyers is independent from the market segment although buyers take more time in the higher priced market segment, brokers who know their (buying) clients taste can reduce the number of visits, however see \cite{Anglin1997}, (vi) higher or more frequent downward LP revisions indicate lower transaction prices and longer TOMs (costly to the seller), properties placed on the market at relatively high LPs and vacant properties are most likely to undergo downward LP revision, while properties with exceptional features are least likely,
see \cite{Knight2002}, range pricing (publishing both SRP, or an even lower price, and LP) dfails to lead to higher transaction prices and it tends to increase TOM, see \cite{AllenFR2005}, and (viii) there is theoretical evidence that a well-functioning rental market helps to equalize hot and cold periods in a housing market, see \cite{Krainer2001}.}
Taken together the extracted reasoning rules may lead to an ASIL  for the market at hand, which a seller may prefer to have available before contemplating any details of getting a selling thread active. 

Suppose that an agent $a$ is involved in decision preparation for a sales thread startup decision on 
$G \in C_{hm}^{80\%}(R)$. Prospective seller $a$  faces the task to make a number of choices concerning pricing: reservation price, list price, etc. Now $a$ may acquire access to a number of research papers $P^i$ dealing with similar markets in regions elsewhere 
($R^i$) and regarding past time intervals ($I_t^i$)  abbreviated as $C_{hm}^{80\%}(R^i,I_t^i)$ for $i \in \{1,...,k\}$.

Each paper $P^i$ yields a family of reasoning rules $F_{rr}^i$ which has been confirmed with 
certainty $v_i \in [0,1]$ according to the authors of $P^i$. For each choice (of a value for say $u$) that $a$ is planning to 
make in preparation of a sales startup decision  the relevance of
both the reasoning rules as well as the market under investigation in paper $P^i$ relative to the issue of determination of an optimal value for the quantity $u$  must be estimated leading to a 
weight factor $w_{i,u} \in [0,1]$. Now in determining an appropriate value for $u$ the reasoning rules $F_{rr}^i$
are applied with weight factor $v_i \cdot w_{i,u}$. 

It should be noticed that these rules may be mutually inconsistent. That may call for the application of some form of paraconsistent logic (see \cite{Middelburg2011} for a survey). At this level of generality the matter may seem rather hopeless but in a practical case, with more structure and contextual information at hand,  it may be manageable. Clearly for specific markets there may be a lot of work that researchers can do in order to facilitate this course of action for individual prospective market participants.

\subsection{Tactical WIA preparations}
Tactical WIA preparations for a selling thread startup decision may include the following:
\subsubsection{Acquisition of selling framework competence}
In the case that $a$ intends to make use of a broker, that is an intermediate agent who carries out most steps in a sales thread for $G$, then it may not be expected in any way that $a$ acquires expert knowledge on the market of $C_e$ goods. In other cases that may be different.  Nevertheless, $a$ will need to play a role in decision preparation for any sales thread startup decision, even if $a$ receives support from a consultant. Some minimal level of knowledge of $a$ on the $C_e$ market must be presupposed, that is a minimal awareness needed to setup an adequate contract with a broker that will act on behalf of $a$. That contract will follow the choices made by $a$ when taking the sales thread startup decision.  In terms of \cite{BDV2011a} $a$ needs 
$C_e$ market framework competence in order to be able to choose an agent $B$ who will provide advice on the parameter settings for a useful sales thread startup decision, probably meant to startup a thread effectuated by $B$.

Classification of ASIL preparation under tactical WIA is plausible because it may be very time consuming while no strategic WIA phase may be present simply because $a$ is not contemplating any other decisions than those in relation to selling $G$. Classification as strategic WIA preparation is less plausible because in that case the assumption is reasonable that the development work is performed as a service to other future decision takers, instead of by a single one just for his own benefit.
\subsubsection{Application of optimizing modifications before putting $G$ on the market}
Once the idea that $G$ will be sold gets momentum, though before the sales thread startup decision is taken, several aspects of its physical condition may need to be taken care of:
\begin{description}
\item{\em Necessary maintenance.} Selling a second hand car with flat tires or broken windows may be impractical, and steps may be needed to arrive a sufficient level of maintenance that is required for an effective sales thread startup decision.
\item{\em Profitable amelioration.} $G$ may look far worse than it might do after being cleaned and upgraded. For instance when selling a home one may want to ged rid of embarrassing old furniture in spite of the fact that it served its owners well for many years.
\item{\em Tactical depersonalisation.} It may be the case that an STS decision is problematic because $G$ is 
carrying too many features that  are personal to $a$. Such features are likely to be removed by every buyer of $G$ and that makes it advisable to remove these personal features before the  sales thread is initiated.
\item{\em External advice on preparatory modification.} Maintenance, amelioration, and depersonalisation each carry additional costs that must me recovered when selling $G$. External advice may be needed to determine which steps are necessary and which steps are probably useful.
\end{description}
Some preparatory modifications may be applied in JIT fashion for the selling thread startup decision or even in JIT fashion for actions rather than implied decisions that will occur during the run of the selling thread once started. An example is cleaning a house which is in full use before it is shown to potential buyers. That action will not be part of 
the JIT preparation for the startup decision.
\subsubsection{Motives for contemplating to  sell}
If $a$ comes to the conclusion that a preference exists for selling good $G$, it is important that $a$ clarifies reasons for that fact because such reasons may be needed to understand the rationale of choices of values, for instance of the reservation prices that $a$ will be committed to. In the OODT perspective decision taking is a means to and end rather than an end in itself. A decision outcome must be gauged against its effectiveness in achieving certain objectives. Being explicit about these objectives as much as possible is necessary for the criterion that the decision taker understands why it is to be expected that effectuation of the decision outcome will have consequences that bring the objectives closer to a realization. Here is a list of possible motives that $a$ may maintain for selling $G$. In each case the motive formulates a problem to the solution of which selling $G$ may contribute. Some problem pairs are mutually exclusive but several of the listed issues can occur concurrently.

Implementing some of these modifications cannot be part of strategic WIA preparation because it is exclusively done in 
preparation of a single future decision. An ASIL that was obtained during strategic WIA preparation may be very helpful, however to determine which modifications are likely to be helpful and cost effective. In that case a classification of optimizing modifications as tactical WIA preparation is convincing. However, some modifications may be applied in JIT mode for the selling thread startup decision. 
\begin{description}
\item{\em Costs too high with limited utility.} The cost of keeping $G$ up and running has become too high, in spite of the fact that $a$ would be keen to make further use of $G$. Even a (moderate) negative price might be effective for removing the burden of the costs of ownership of $G$. A typical example is that $G$ is a car: the cost of ownership disappear instantly together with the ownership. Paying for being released from the ownership may be justified if no ordinary buyer can be found. The same may hold for a yacht. A dwelling for recreational purposes may force its owner to pay  high annual fees to the operator of the park where the site is located. Getting rid of that obligation may be quite difficult and paying for it may be the only option.
\item{\em Costs too high in spite of high utility.} The need for $G$ or for a cheaper instance of class $C_e$ may be high, but the cost of ownership for $G$ may have become too high for $a$, either because of changes in $a$'s circumstances (lower income or financial mishaps) or because of an increase in the cost of ownership of $G$,  so that $a$ develops a preference for replacing $G$ by a simpler and cheaper instance of the same class.
\item{\em Utility too low.} If the  utility of being able to make use of the ownership of $G$ has degraded too much, there is no point to continuation of its ownership, even if the cost of it are bearable. A typical example is a cottage that is not used anymore because of changed interests or because after moving to another home the distance to its site has become impractical. Another example is that $a$ has developed a negative emotional attitude towards $G$ and for that reason prefers to avoid making use of it.
\item{\em Expected degradation of utility.} Rather than that utility has become too low, that may be expected in the near future and selling $G$ is contemplated to prevent an unwanted situation.
\item{\em Utility has degraded and invested funds must be recovered.} If an owner occupier of a home gets a new job elsewhere and commuting is not an option then not only has the utility of the home degraded but selling it is also an important step to recover the funds that have been invested in it. These funds will be needed to arrange housing near the new job.
\item{\em Risk of degradation of value.} If $G$ constitutes an important part of what $a$ owns, and if $a$ fears that the value of $G$ may decrease for external reasons out of $a$'s control, $a$ may develop a preference for not owning $G$ anymore. For instance, if $a$ is the owner occupier of $G$ then  $a$ may fear that a new building site will be developed nearby thus making $G$ far less attractive for new buyer in the region. Selling soon may produce a better price than selling later. Selling soon, and then renting and buying a home in the new development may be the most promising method for protecting $a$'s wealth.
\item{\em Realization of expected profit.} If $a$ presumes the marketvalue of $G$ to be quite high so that when selling $G$ making a substantial profit is likely, that may constitute a motive for preparing a sales thread startup decision. 
\item{\em Utility has become too low given available financial means.} If $a$ is able to replace $G$ by another instance $G^{\prime}$ of the class $C_e$ of higher quality and higher cost, then selling $G$ and using the 
financial means thus obtained for buying $G^{\prime}$ may constitute a step forward for $a$. In such cases $a$ may be uncertain about the price that will be made for $G$ so that sell first and buy later is used as a safe way of going ahead.
\end{description}

It is plausible that some weighted combination of these motives constitutes the ``real motive''. In that case its is meaningful to determine which motives contribute to $a$'s preference for selling $G$ and to what extent.

\subsection{JIT preparation 1: determination of 5 prices and 2 values}
Before taking the selling thread startup decision a number of quantities (prices as well as values) must be determined.
Because of fluctuation in time strategic WIA preparation cannot provide these quantities in a reliable fashion and 
JIT preparation is a more plausible classification.

Some of these values are needed for decision taking while the selling thread is active. Thus some of the JIT preparations for the selling thread startup decision are at the same time tactical WIA preparations for subsequent expected selling thread related decisions, with the bid acceptance decision as a key example because it may be necessary to  take a bid acceptance decision, or a bid rejection decision under time pressure, leaving no time for JIT preparations.

\subsubsection{Inner circle separating reservation price}
Suppose $a$ sells $G$ for a very low price $P$ to an unknown agent $b$. Then $a$ may be blamed by his friends or relatives for not offering them the opportunity to acquire the ownership of $G$ for a comparable, or even higher price. Thus in a low price range $a$ may maintain a list of preferred potential buyers.
A safeguard against unintended preferred buyer miss is needed. 

This safeguard is found if some price $Q_{ic}$ can be determined such that (i) no 
member of IC$_a$ is able, or willing,  to pay $Q_{ic}$ for obtaining the ownership of $G$, and (ii) $a$ will commit not to sell $G$ at or below $Q_{ic}$ to a non-preferreed buyer (that is buyer outside IC$_a$). Such a price will be called an inner circle separating reservation price (ICSRP). ICSRP separates the inner circle from an external world where a buyer will be sought by running a selling thread. Unavoidably ICSRP depends on the time $r$ of its assessment, which is indicated by ICSRP$_r$, without a subscript ICSRP denotes the current value.

The ICSRP  that $a$ attaches to $G$ is mainly influenced by two aspects: size (scope) of IC$_a$, and the 
financial strength of the members of IC$_a$, (assuming that the ICSRP lies well below the market value of $G$). In many cases, and in particular if IC$_a$ consists only of close relatives of $a$ 
a useful value of $Q_{ic}$ may be found without any careful inspection of $G$ because the members of IC$_a$ are far from being able to pay a plausible market value for $G$.

When a decision to sell $G$ is to be taken at time $r$ by $a$ a value for ICSRP$_r$ must be determined for inclusion in the decision outcome. If such a price has not been found in advance subsequent selling is sensitive to selling against too low a price, thus leading to preferred buyer miss. 

If one makes use of willingness to pay as a predicate over agents, objects, amounts of money and time it is possible to
define $Q_{ic}$ uniquely as  the maximum of the prices that any of the preferred buyers is willing to pay fro $G$ during TOM after selling thread startup. 

\subsubsection{Final seller reservation price, seller reservation time, and seller reservation certainty}
As components included in a decision outcome concerning the sale of $G$ it is necessary that $a$ formulates three logically (or economically) connected quantities:
\begin{enumerate}
\item   (seller) reservation price (FSRP), that is a minimum price below which $a$ will not sell $G$, and at which $a$ is willing to sell after expiration of SRT (see below),
\item  (seller) reservation time (SRT), that is a maximum to the TOM (time on market, that is the time that $a$ allows $G$ to be on the market until selling or until giving up), and
\item (seller) reservation certainty (SRC), that is a lower bound for the estimated probability that the seller $a$ will 
complete a sale of $G$ against a price at or above SRP with a TOM  at or below SRT. 
\end{enumerate}
Obviously FSRP, SRT, and SRC must be chosen simultaneously and with acknowledgment of their interrelations.

As a simplification we may assume a default value for SRC at 75\%. Then it will be required that the probability that $G$ is sold for a price at or above FSRP (given prices settings and acceptance protocol as specified in the selling thread startup decision) and with TOM below SRT must be estimated at least at 75$\%$.  FSRP depends on $t=$ SRT, on the absolute time $r$ at which the selling thread is supposed to begin, which for simplicity is identified with the moment of startup decision taking, and on the $a$'s private situation (which may be unknown to other market participants), and this dependency is made explicit by writing SRP$_{t,r,a}$ below.

When taking the selling thread startup decision it may be assumed that ICSRP$_r$ $<$ FSRP$_{t,r,a}$. Both quantifies are fluctuating in time and as a consequence their order may change.
If during decision preparation $a$ finds out that say FSRP$_{t,r,a}$ $\leq$ ICSRP$_r$ it is plausible consider whether or not to avoid starting up a selling thread altogether and to look for a buyer within $a$'s inner circle IC$_a$ perhaps without making use of a broker. During the run of a selling tread new insights concerning FSRP and ICSRP
may surface at time say $s$ after thread initialization  to the extent that  
FSRP$_{t-s,r+s,a}$ $\leq$ ICSRP$_{r+s}$ which may indicate that taking a decision for 
terminating the selling tread is plausible, or at least needs to be considered, in particular if a large proportion of SRT has already elapsed. 

SRP$_{t,r,a}$ $\leq$ ICSRP$_r$ may turn out to be the case because of the default setting of SRC at 75\% this can be remedied if $a$ is willing to start a selling thread with a lower value $p$ for SRC. In this way $p$ enters the determination of FSRP thus suggesting the notation FSRP$_{t,r,a,p}$.%
\footnote{One may take a market average of TOM's for successful selling threads as a default value for SRT and then incorporate both $a$'s willingness to accept a longer SRT as well as a need to require a shorter SRT in the conditions that come along with $a$'s private circumstances. A short SRT  seems relevant only if a high 
SRC is aimed for, thus lowering FSRP, while a long SRT that may go hand in hand with a relatively low SRC suggests that SRP can be increased. It should be noticed that by lowering SRC or by increasing SRT, the value of FSRP can be made arbitrarily high. Thus fixing FSRP without a simultaneous awareness of STR and SRC is rather meaningless.}
Now with $p <$ 75\%, ICSRP$_r$ $<$ FSRP$_{t,r,a,p}$ may be recovered thus preventing this aspect to frustrate a selling thread startup decision to be taken.

For the sequel of this paper I will assume that there is no functional dependency between FSRP and $p$, thus writing
FSRP$_{t,r,a}$, which indicates that $t$, $r$ and the private circumstances of $a$ matter immediately for determining the FSRP.

SRP$_{t,r,a}$ may be determined in many different ways depending on $a$'s situation. For instance when selling a home it may be known which home (say $G_n$) $a$ will buy once selling $G$ has succeeded, and for what price the new home can be obtained. Now $a$ may calculate what selling price for $G$ he needs to acquire $G$ subsequently and to arrive in a financially acceptable state. If $a$ has already conditionally bought $G_n$ with the proviso that $a$ can be sold at some given price, $a$ may choose to sell for a lower price if that fit the long or medium term financial objectives of $s$.%
\footnote{It follows that in the housing market, by conditionally buying a new home, a seller may find a way to compute SRP in a satisfactory way.}

\subsubsection{Subjective market value}
SMV$_{t,r,a}$ is the price that $a$ (or rather a number of experts watching $a$) expects to 
make by selling $G$ within the time $t=$ SRT with probability 
(certainty) $p$, starting the selling activity at time $r$ given the particular considerations that $a$ may have. Clearly it
is plausible that when taking a selling thread startup decision one finds SRP$_{t,r,a,p}$ $<$ SMV$_{t,r,a,p}$.
 
 If when JIT preparing for a selling thread startup decision at time $r$ (also considered to be the 
 approximate time of the decision ahead), $a$ notices that 
If SRP$_{t,r,a}$ $\geq$ SMV$_{t,r,a,p}$ then some mistake must have been made in the determination of SRP, SRT and SRC, and no startup decision can be taken with exactly those figures, thus necessitating a revision of that part of the JIP preparation.

\subsubsection{Initial selling reservation price}
ISRP, the initial minimal selling price, at or above the FSRP, but at or below SMV is a price below which $a$ will not sell initially. A plausible stopping criterion is to interpolate linearly between ISRP and SRP so that offers made by a potential buyer at time $r+$TOM which are at or  in excess of the interpolated intermediate reservation price
FSRP + (SRT - TOM)/SRT $\cdot$  (ISRP - FSRP) will trigger a positive bid acceptance decision by $a$.

\subsubsection{Objective market value}
An objective market value, or simply market value (MV$_{r}$), to be obtained within the objectively expected TOM time (OETOM, not SRT which depends on $a$ whereas OETOM is supposed to be valid in general) starting at absolute time $r$  for $G$, needs to be formulated as a part of a selling thread startup decision for $G$ by its owner $a$.%
\footnote{MV$_{\texttt{now}}$    is also called the current MV.}
 Here objectivity refers to the idea that  MV$_{r}$ is the price expected to be obtained by a normal seller who has no particular constraints and is not under any specific pressure to sell.%
 \footnote{The notion of a normal seller is rather hypothetical. In this case marketing and selling are identified, the difference between both functions being rather a matter for large volume markets. Further one must assume average selling competence rather than optimal selling competence, even if all sellers will try to engage a broker who operate at or above the average competence level.} 
MV$_{r}$ assumes that a successful sale is to be obtained with a probability of say $75\%$ (this figure must also be gauged against market practices in a specific market).%
\footnote{Thus while SRP depends on $a$ (or rather on the current status of $a$), as well as on the SRT, the current MV given SRT does not depend on $a$.}

Credible determination of MV$_{r}$ may be quite difficult%
\footnote{If $G$ is unique, that is if there are relatively few objects comparable to $G$ then determination of a 
plausible current value for MV given SRT may be an almost impossible task. Further, if a broker, say $B$ mediates between $a$ and the market and the broker suggests values of current MV and SRT ending up in a contract between $a$ and $B$ then it may be reasonable to include a penalty for $B$ if no offer at or above the estimated current MV  has been observed before the expiration of SRT after announcing that $G$ is for sale $G$ on the relevant market. Of course this rule makes sense only if $B$ cannot trigger the occurrence of fake bids.}
and for several entity classes it constitutes a classical field of professional activity underpinned with significant volumes of research, and often regulated by formal bodies.%
\footnote{To mention two papers from a vast volume of work: in \cite{AlukoAA2004} the case is made that valuation of land should aim for prediction of future value rather than mere interpolation on the basis of recent market data; in \cite{Goldberg1982} (fair) market value is compared with replacement value and it is found that this comparison depends on circumstantial facts so that preferences for one or the other are not uniformly valid in al market conditions.}
Some remarks can be made here:
\begin{enumerate}
\item  If nearby objects quite comparable to $G$ (in a hedonic valuation model) are on the market for prices 
below $Q$ and for a period exceeding SRT  then it is plausible to consider $Q$ an upper bound for 
MV$_{r}$. 
\item If a nearby object which outperforms $G$ in a hedonic valuation model has recently been sold for $Q$ then that constitutes a plausible upper bound for MV$_{r}$.
\item If a nearby object which underperforms $G$ in a hedonic valuation model has recently been sold for $Q$ then that constitutes a plausible lower bound for MV$_{r}$.
\item If ample recent data about TOM and transaction price  are available for comparable 
objects averaging those figures may provide a plausible estimate bound for MV$_r$. 
\item If general trends are known concerning the transaction prices and TOMs of objects comparable to $G$ then extrapolation from older and scarce data on transactions of objects comparable with $G$ may be used to estimate its current MV, or conversely. 
\end{enumerate}
It is reasonable to require that SMV$_{\texttt{SRT},r,a}$ $\leq$ MV$_{r}$ if SRT $<$ OETOM and conversely if $a$ has less time pressure  than the market average (SRT $>$ OETOM), it is plausible that 
SMV$_{\texttt{SRT},r,a}$ $\geq$ MV$_{r}$. 

However, if SMT = OETOM and at the same time  SMV$_{\texttt{SRT},r,a}$ $>$ MV$_{r}$, there is a problem which must be addressed before taking the selling thread startup decision because this state of affairs is implausible.  Indeed 
it is important that at the time of taking  a selling thread startup 
decision $a$ is aware that SMV$_{t,r,a}$ $\neq$ MV$_{r}$ if that happens to be the case.

\subsubsection{List price}
Assuming that after selling thread startup decision for $G$ has been taken by $a$ information derived from the corresponding decision outcome will be placed on one or more lists, as a means to make its being on the market publicly known, then a list price (LP) must be chosen as the main figure which will be made public. 

Under the assumptions made in Paragraph \ref{Constr} that offers are expected  not in excess of LP it is reasonable to require that 
MV$_{r,a}$ $\leq$ LP$_{t,r}$. Apart from that the determination of a list price is a matter of selling strategy. Downwards adaptation of a list price can be applied while a selling tread is running. Downwards adaptation of ICSRP, SRP,
and SMV is much less plausible. Only having determined an LP a seller is unprepared for implied bid acceptance decisions. In the setting outlined above LP plays no role for taking bid acceptance decisions.%
\footnote{In some markets, for instance the Dutch housing market, the list price dominates all other prices and values. That seems to be an ineffective method which allows sellers to stay unaware of unpleasant aspects of a market much longer than needed.}

A list price is a publicly available  asking price open for (upward or downward, depending on conventions) negotiation. List prices are common for realtors (state agents) who announce their offerings on lists and who often maintain a MLS (multiple listing service) in combination with a group of colleagues.%
\footnote{In the Netherlands the famous site \texttt{www.funda.nl} is an example of an online MLS for the housing market.}

\subsubsection{Ideal price}
The owner $a$ may have in mind a price that he would hope for in good times. This is the ideal price (IP$_r$, at the time of decision taking). It is plausible to assume that MV$_{r,a}$ $<$ IP$_r$. Having obtained IP$_r$ 
for $G$ within SRT from $r$ would be most satisfactory for $a$. If that comes true there will be probably 
no negative afterthoughts by $a$ or members of his inner circle about the sale. Instead such feelings may then hit the seller. A reasonable default for IP, useful if one prefers to ignore IP,  it is to equate IP with LP.

\subsection{JIT preparation 2: bubble/burst detection and risk analysis}
Finally several aspects need to be taken into account, however briefly: bubbles, bursts and other risks.
\subsubsection{Bubble detection and burst accommodation}
The algorithm used to take bid acceptance decisions is complicated by two factors: (i) there may be a bubble in the market and more prospects may seek information about $G$ than either seller or broker expected. In that case either LP must be quickly increased with the potential collateral damage of rendering expensive marketing materials containing the original LP redundant, or potential buyers must be informed that offers above LP are most welcome and 
that selling above LP is expected by $a$, (ii) there may be a burst in the market with as a consequence that the market proceeds so slowly that no prospects take noticeable action about the fact that $G$ is for sale. In this second case there is no other option, except forgetting about the sale altogether, than either to lower LP or to increase marketing efforts and cost (which amounts to lowering SRP, an step with even more impact). When to engage in either one of these options (f any) is a difficult optimization problem in each individual case. 

Preparation for this event is possible if during JIT preparations not only SRP, SRT, and SRC but also SRPF (seller reservation prospect frequency, measured in new prospects who entered the process as potential buyers per day) has been determined, where as soon as SRPF $\cdot$ TOM exceeds the number of unique prospects that has been registered by the broker  as having been active until that moment, this signal (condition)  is used as a warning that LP must be decreased.

The risk of missing a bubble is similar to the risk of underpricing (LP too low) and for that risk the remedy is the same as for bubble missing. Symmetrically the risk of missing a burst is similar to the risk of overpricing with the same strategy being applicable in case of overpricing.

\subsubsection{Risk analysis}
OODT style decision preparation for a selling thread startup decision (for a good $G$ of type $C_e$) as suggested above need not be the best way of going ahead for a potential seller $a$. Risks implicit in the approach need to be formulated and analyzed. Here are some plausible risks that need to be addressed:
\begin{enumerate}
\item Too much focus on estimating prices like ICSRP, SRP, SESP, OESP, LP and IP, in a case where prices are primarily produced by individual transactions.
\item Lack of flexibility when the ordering between the various prices (ICSRP etc.) fluctuates during the active period of a selling thread.
\item Risk of preferred buyer miss, for instance because the willingness to pay of some preferred buyers increases (unnoticed by $a$ and $B$) during the run of the selling thread.
\item Risk that buyers who may offer unexpectedly high prices are ignored or are not given an incentive to do so.
\item Lack of flexibility when market values fluctuate quickly especially if low margins over the reservation price are to be expected.
\item Lack of focus on the qualities and working methods of the broker hired by the seller,%
\footnote{For instance: is adaptive selling needed (See \cite{SpiroWeitz1990})? How is the broker's sales competence measured, and is he learning on a permanent basis (see \cite{ChonkoDJR2003})? Is the broker making use of an acknowledged set of ``best practices'' (see \cite{LeighMarshall2001}?)}
 and underestimation of what needs to be done in order to determine OESP.
\item Unawareness of the existence of fundamentally different channels for selling goods of type $C_e$. Implicit choice for a channel through broker selection.
\item Inability to allow ``intuition'' to play a decisive role when that might just work.
\item Difficulty to imagine from which perspectives, in what periods, and by whom, the decision outcome will be assessed, judged, or otherwise evaluated.%
\footnote{
A decision to acquire an insurance policy against a certain risk may be assessed after its expiration date with the additional information that the risk did not materialize. That kind of meta-risk has been underestimated by several Dutch educational institutions who bought interest swaps several years ago to insure against the risk of increasing interest rates now to find out that not only interest rates went down thereby turning the swaps into a very costly obligation rather than a useful possession, but also to find out in some cases that the decision to buy swaps is now being portrayed as a ``speculation with public money'' by some unfriendly but audible commentators.}
\item A broker acting on behalf of the seller may be vulnerable for so-called paradoxes of trust (see \cite{Oakes1990}). The prospective seller must ensure not to engage in too simple assumptions concerning broker behavior.
\item Psychological risk for $a$'s reasoning methods, for instance (i) risk of cognitive inertia (see \cite{Hodgkinson1997}), as a consequence of focusing too much on self-determined values for various prices, (ii) one of the flawed arguments (so-called traps) listed in \cite{HammondKR1998}, and inaction inertia (see \cite{ButlerHighhouse2000}).
\item Methodological problems: for instance,  one may use Bayesian update theory to develop an estimate of prices of $C_e$ goods, given a known and confirmed but slightly outdated price distribution for computing prior odds and scattered recent sales figures for which to derive updates. But market prices depend on the behavior of other (potential) buyers and according to \cite{Camerer1998} these may not use Bayesian updates but rather some other update rule, thus rendering the predictive value of Bayesian updates for the given seller $a$ problematic.
\item Performance problems: for instance in the health sector a bibliography on informed decision making is so large that acquiring an integrated view on the subject may be prohibitive (see \cite{BekkerEA1999}). How to become informed on informed decision making as a prospective seller in a particular market.
\end{enumerate}

Risk analysis has two sides. It may lead to reconsideration of parts of decision preparation so that the candidate decision outcome is modified. This suggests that risk analysis should be applied repeatedly, until its impact on the decision outcome under preparation has vanished. Risk analysis may also be performed (leading to a statement termed `` risk analysis'')  in a final stage of decision preparation leading to a report which indicates that the proposed candidate decision outcome is defensible against a family of risks, understood as possible future ``attacks'' on the decision outcome after the decision has been taken in correspondence with the preparations.

\subsection{JIT preparation 3: broker selection}
Selecting a broker, or self-appointment in case no broker will be hired and  the broker role is played by $a$. This is an unavoidable part of decision preparation for selling thread startup, which can be done in JIT mode, or in tactical WIA mode or, if a long term association with the same broker for a number of similar but different transactions has been planned broker selection belongs to strategic WIA preparation.

\section{Joint decisions with multiple decision outcomes}
\label{JDec}
As proposed in Section \ref{TTfD} a classification of decisions can be done through decision types. 
A decision type consists of two or three parts: (i) a
 so-called DOT (decision outcome type), (ii) a decision taking protocol (DTP) which indicates how a decision is taken after preparations have been made, and optionally (iii) a decision preparation protocol (DPP) which indicates how decision preparation must be performed. Below only a DOT will be specified, leaving the development of the corresponding DTP and DPP to the reader's imagination.

It is unfeasible to provide an analysis of each of the selling decisions in this paper. Two types of decision stand out, however: the sales thread startup decision and the sales transaction decision. In both cases several modalities must be distinguished.
\begin{description}
\item{\em Owner $a$ takes single actor decision.} A sales transaction decision can be a single actor decision if a binding accept of an offer from a potential buyer can be made by the (prospective) seller. 

A sales thread startup decision for $G$ can be a single actor decision taken by its owner $a$ if $a$ is confident that either no mediating agent (broker) will be needed or that $a$ will definitely find an intermediate agent who is willing to work ion the basis of the given decision outcome.
\item{\em Owner (prospective seller) $a$ takes joint actor decision with broker $B$.} This case applies to the sales thread startup decision.

Decision preparation and decision taking are shared by $a$ and $B$. A joint decision outcome is produced.
\item{\em Owner (seller) $a$ takes joint actor decision with buyer $b$.} This case applies to the sales transaction decision.

Decision preparation and decision taking are shared by $a$ and $b$. A joint decision outcome is produced.
\end{description}

\subsection{Single actor selling thread startup decision with embedded broker proposal}
\label{DOTsts}
The default option, and indeed the simplest case is that a broker will be engaged and that the selling decision at the same time is a binding proposal to a broker asking him to ``run'' the selling thread on behalf of the seller. In that case it is plausible, though not necessary, that the decision outcome has been prepared with that broker as a consultant, so that an abstraction of the decision outcome serves as an offer to the broker for making use of his service.

A single actor selling thread startup decision by $a$ about $G$ must produce a decision outcome that
provides at least the following information.

\begin{description}
\item{\em Object presentation.} It needs to be stated in words and pictures what $G$ amounts to, so that by providing the object presentation no doubt can arise as to what is for sale.
\item{\em Price settings.} A numerical value is chosen for each of the following parameters.
\begin{enumerate}
\item Inner circle separating reservation price (unless sale to a member of the inner circle is intended),
\item seller reservation time,
\item seller reservation price%
\footnote{Taking expected broker commission and listing service cost into account.}
\item subjective market value,
\item market value,
\item list price (usually above the expected selling price, assuming that bidding below list price is the relevant market convention for goods like $G$),
\item seller side ideal selling price (what the seller hopes for; above the market value, sometimes also above the list price).
\end{enumerate}

\item{\em Broker data.} A proposed broker is mentioned together with his business data which uniquely identify his company/person. It is assumed that $a$ expects the broker to agree with the assignment as well as with the various parameter settings.%
\footnote{The seller reservation price is confidential information, and the broker may not disclose that price to other market participants in any case, in particular not if he chooses not to accept the assignment.}
\item{\em Marketing method.} Marketing threads will be started for one or more media (for simplicity all media are referred to as listing services). Some (explicitly named) listings may be activated without broker involvement, other listings may be subsequently be activated by the broker in the course of selling thread effectuation.

\item{\em Reasons for the selling intention.} Awareness of motives is needed for exception handling, which may be needed when running the selling thread.%
\footnote{In each of the following circumstances, $a$ may be asked to produce an implied decision and  having an unambiguous awareness of the motives for selling may be a critical asset for seller and broker alike.
\begin{enumerate}
\item dealing with offers below the seller reservation price, 
\item rejecting offers above the reservation price,
\item waiting until a reasonable approximation of the market value has been offered, 
\item terminating a marketing thread (before termination of the parent selling thread),
\item terminating a selling thread (before a bid has been accepted).
\end{enumerate}
}
 
\item{\em Viewpoint on market dynamics.} Prospective seller $a$ must be aware of potential 
bubbles and bursts in the $C_e$ market and express some opinion on the current market, 
as well as the implications thereof for the effectuation of sales threads.
\end{description}

\subsection{Fragmentation of the selling thread startup decision outcome}
The outcome of a selling thread startup decision, viewed as a of piece of information is split into parts each constituting an abstraction of the decision outcome obtained by deleting a part of the data, and each meant for different audiences:
\begin{description}
\item{\em Self.} Prospective seller $s$ receives the full decision outcome.
\item{\em Inner circle.} The inner circle receives the following abstraction of the decision outcome: object presentation (probably abbreviated, depending on what members of the inner circle are expected to know about $G$ already), inner circle separating reservation price, broker identity (if present), information on marketing methods, and reasons for the selling intention.%
\footnote{ Members of the inner circle must be made aware that is will be an assumption for the seller that the will not even contemplate making offers at or above the inner circle separating reservation price. If that turns out to be wrong the selling thread must be aborted and may be restarted with a higher inner circle separating reservation price only, but  a private sale to a member of the inner circle may result as well without a second selling thread having been started.}
\item{\em Broker.} A broker obtains: seller reservation price and time, expected broker commission, expected selling price, list price, marketing method, stopping criterion, and viewpoint on market dynamics,
\item{\em Listing services.} Listings (placements on a listing service) may be effected by the seller directly or via the broker's mediation. In both cases besides technical information about $G$ listing services are provided a list price.
\end{description}
Upon taking a selling thread startup decision the decision outcome is split into a family of multiple outcomes each directed to its own audience as specified above.

\subsection{Two variations on the  theme of a DOT for selling thread startup decisions}
At least two variations on the theme of a single actor selling thread startup decision with embedded broker proposal must be considered.
\subsubsection{The case without a broker} The case that no broker is proposed is difficult because it hardly 
complies with the understanding of a decision from \cite{Bergstra2011a}. The matter is that the very coming into existence of a decision outcome (including its distribution to the intended audience) must initiate a chain of events that eventually brings the objectives of the decision taker closer to being realized in a way predicted by the decision taker. 

The only ``solution'' to this difficulty is that immediately after having taken the decision $a$ distinguishes two roles, seller and broker, and plays both roles concurrently and independently, while broker commission is set to zero. In the broker role $a$ may not modify the parameter settings of the decision outcome unless in the role of an owner $a$ has taken corresponding decisions. 

This seems to be an artificial matter but the setup disallows seller $a$ dropping data from the decision outcome that he must transfer to himself in the other role. Preferably the decision outcome is stored with an independent agent so that the appropriateness of the course of events can be judged afterwards.%
\footnote{A similar case of role splitting occurs when a unit $U$ is preparing the outsourcing of sourcement $s$ in the terminology of \cite{BDV2011b}. Before performing the outsourcing step $U$ is often advised to proceed through a phase of explicit demand management in which $U$ plays simultaneously the roles of provider and of a client of $s$, thus learning what it means to act as a client (that is perform demand management) before forcing itself into that role (by outsourcing).}
\subsubsection{Taking a joint actor selling thread startup decision} This case is very similar to the single actor selling thread startup decision with embedded broker proposal with the difference that: (i) a broker (say $B$) is engaged with certainty as side-effect of taking the decision, (ii) $B$ has been consulted in advance (that is, has participated in decision preparation) to the extent
that he can and will take the decision to be engaged as a broker in the selling thread at the same time by way of a joint decision with $a$, (iii) non-broker mediated listings services need not sent abstractions of the selling decision (so that they will perform the envisaged listing) and all listing activities may be postponed as broker actions.

\subsubsection{Commentary}
One may criticize the decision types just specified for being far too complex. However, I hold that for instance  in the case of selling a home the  these requirements on decision outcomes are not exaggerated. All information will wrapped as parameters in the specified decision outcome be needed if effectuating the selling thread is to result in a valid selling process.

An artificial aspect is the case without a broker where the owner is supposed to create a second role (as a broker) for himself while deciding to startup a selling trace. But this is not overdone: maintaining a separation between what is to be achieved and under which constraints (all specified in the decision outcome) from how that is achieved and by whom that is to be done will prove meaningful for the owner/seller.

\section{Concluding remarks}
An application of OODT to the description of selling threads and selling thread startup decisions has been outlined in considerable detail. Three aspects deserve being highlighted by way of conclusion: (i) what use can this development potentially have for selling agents and their brokers, (ii) I will mention 
cascading decision taking and cascading decision types as a possible application ready for further research, and (iii)
how is the relation between the terminology of OODT and the concept of thread startup as used for instruction sequence effectuation in the approach of  \cite{Bergstra2011,BergstraMiddelburg2012} and \cite{BergstraMiddelburg2012b}.

\subsection{Potential benefits of OODT for selling agents and their brokers}
\label{PotDel}
The specification of decision taking for selling agents, and in particular for selling thread startup as it is outlined below may be considered to constitute an OODT style theory of decision taking for selling agents focused on the initializing decision. When required to provide a motivation for this work in terms of potential applications I will rely on the notion of a conjectural ability following \cite{BDV2011a}. The idea is that by getting acquainted to this  ``theory'' a person with framework competence on serving as a seller or as a broker, will develop a number of so-called conjectural abilities.
These conjectural abilities are so-called potential deliverables or benefits  expected from 
(becoming acquainted with)  the theory.

Here are some deliverables that may sprout from an OODT style investigation of decision taking in a specific context or market.
\begin{description}
\item {\em Protocol design for selling thread startup decision taking.} For decision taking that leads to the initialization of a selling thread protocols may need to be worked out in advance to make sure that no important steps are forgotten.%
\footnote{Instead of the term ``protocol''  the phrase ``instruction sequence'' might have been used as well. Protocols often are used as definitions for the functions that can be computed by their effectuation, that is such functions are laid down by how they are computed rather than by means of giving an abstract input-output specification.}

\item{\em  Thread mining.} The conceptual structure in terms of threads may support the mining of protocols or threads used by selling agents and their brokers in practice. That form of mining may be helpful for protocol design.

\item{\em Thread suggestion for a community.} If decision taking, as practiced by selling market participants in a certain market can be identified as a cause of poor performance of that market it constitutes a conjectural ability to ``discover'' new decision taking threads (protocols) and to communicate these as operational options, or even suggestions,  to a large group of current or future market participants, with the intention to improve the working of that market in a structural manner.

\item{\em Thread suggestion for an individual market participant.} Of course one may simply intend to strengthen the operations of a single market participant by suggesting it to make use of novel threads involving decision taking. The ability to do so on the basis of the ``theory'' at hand is conjectural, at this stage.

\item {\em Decision taking support.} The design of automated support for decision taking requires some 
model of decision taking threads. To compare with software engineering, a software engineering workbench may be based on a specific life-cycle model even if adherence to that particular life-cycle by teams in the software engineering community  has not yet been observed in software engineering practice. A successful tool may generate its own practice.

\item{\em Artificial decision taking.} It was  claimed in \cite{Bergstra2011a,Bergstra2012a,Bergstra2012b}
that decision taking is most plausible for human agents, implausible for animals,  and less plausible for artificial agents. nevertheless the ambition to automate decision taking is an obvious one and its realization may be supported by the development of models in terms of of decision taking threads.

\end{description}
I will not assume that  market participants, and in particular sellers and brokers who are the primary agents that I will discuss below, are likely to be aware of the kind of theory that is outlined or that these market participants operate in any way consistently with the patterns that I am suggesting. 

\subsection{Cascading decision types}
In Paragraph \ref{Plans} the possible roles of plans and plan taking  in guiding the decision making process has been outlined. It is reasonable to replace the plan by a thread, a so-called decision making thread, and to assume that a startup decision is taken to activate the decision making thread. The startup decision for a decision making thread (including JIT preparation and tactical WIA preparation) designed for taking a decision of type $D$ is a so-called cascading decision, and the type $D_C(D)$  for a decision of that type is a cascading decision type.

Cascading decision types are relevant if a decision making process is so complex, risky, or expensive that it preferably comes about from taking a decision to that extent. That decision will trigger the effectuation of an instruction sequence thus producing a thread that implements (or simply represents) a useful decision making process. Selling thread 
startup decisions may be preferably made by means of a  thread started up by a cascading decision to that extent if the good that is for sale is very expensive and or very unusual and if also 
complex and expensive preparations need to be taken.

\subsection{Computer instruction sequence effectuation threads}
Finally  I will provide an example of how the terminology that has been outlined in this paper for OODT and the terminology of threads and instruction sequence execution may be used in combination, thus indicating that gross inconsistencies between both uses of the terminology have been successfully avoided.

I assume the existence of a class IS$_{mac-k}$ of instruction sequences, each of which can be effectuated on a machine of machine architecture class $k$. If the occurrence of a machine operation starts the effectuation of $x \in $ IS$_{mac-k}$ on machine $M$ (of machine architecture class $k$) then this machine operation  may in some cases be said to perform a choice for doing so but following \cite{Bergstra2012b} it is not plausible to hold that $M$ or an operating system running on $M$ decides to start a thread for effectuating $x$ on $M$. Postulating an intermediate role of a decision outcome is  implausible for a machine operation.

A simpler situation pertains if each effectuation of instruction sequences is brought about by the actions of a human operator. In that case the activity of the human operator is probably not an instance of decision taking, but rather an instance of action determination as discussed in \cite{Bergstra2012b}. Assuming that a human operator must determine to start a thread for putting $x$ into effect, and that this works the same for all other instruction sequences from the class IS$_{mac-k}$, 
and assuming that the halting problem for IS$_{mac-k}$ on $M$ is autosolvable (see \cite{BergstraMiddelburg2012}),
then a human operator  may employ a protocol for action determination that first solves the halting problem for $x$ given its input (which must be extracted from the sate of $M$ at the moment of effectuating the halting solver, under the assumption that no side effect of the halting solver, when having come to a positive conclusion, modifies that state
to an extent which matters for a subsequent effectuation of $x$) and then asks for a thread putting $x$ into effect only in the case of a positive result (that is if halting of the effectuation of $x$ is expected).

Now, if there is a group $G$ of operators and if $a$ plays a leading role in $G$, then it is plausible that $a$ 
decides that all members of $G$ are expected make use of the mentioned policy when determining if and how to put an 
$x \in $ IS$_{mac-k}$ into effect on $M$. Indeed when addressing all members of $G$ with this intention, making use of  the mechanism of an OODT style  decision, the outcome of which is broadcasted to $G$ makes perfect sense.


\begin{thebibliography}{99}

\bibitem{AllenFR2005}
M.T.\ Allen, S.\ Faircloth, and R.C.\ Rutherford.
\newblock The impact of range pricing on marketing time and transaction price: 
a better mousetrap for the home market?
\newblock {\em Journal of Real Estate Finance and Economics}, Vol. 31 (1) pp. 71--82, (2005).

\bibitem{AlukoAA2004}
B.T.\ Aluko, C.A.\ Ajayl, and A.-R.\ Amidu.
\newblock The estate surveyors and valuers and the magic number: a point estimate or a range of value?
\newblock {\em International Journal of Strategic Property Management}, 8 (3), pp. 149--162 (2007).

\bibitem{Anglin1997}
P.M.\ Anglin.
\newblock Determinants of buyer search in a housing market.
\newblock {\em Real Estate Economics}, Vol. 25 (4) pp. 567--589, (1997).

\bibitem{AnglinRS2003}
P.M.\ Anglin, R.\ Rutherford, and T.M.\ Springer.
\newblock The trade-off between the selling price of residential properties an the time-on-the-market:
the impact of price setting.
\newblock {\em Journal of Real Estate Finance and Economics}, Vol. 26 (1) pp. 95--111, (2003).

\bibitem{BaetenBastenReniers2009}
J.C.M.\ Baeten, T.\ Basten, and M.A.\ Reniers.
\newblock Process algebra: equational theories of communicating processes.
\newblock {\em Cambridge Tracts in Theoretical Computer Science.} Vol. 50, (2009).

\bibitem{BekkerEA1999}
H.\ Bekker et. al.
\newblock Informed decision making: an annotated bibliography and systematic review. 
\newblock {\em Health Technology Assessment}, 3 (1), (1999).

\bibitem{Bergstra2010}
J.A.\ Bergstra.
\newblock Steering fragments of instruction sequences.
\newblock {\tt arXiv:1010.2850 [cs.PL]}, (2010).

\bibitem{Bergstra2011}
J.A.\ Bergstra.
\newblock Putting instruction sequences into effect.
\newblock {\tt arXiv:1110.1866 [cs.PL]}, (2011).

\bibitem{Bergstra2011a}
J.A.\ Bergstra.
\newblock Informatics perspectives on decision taking.
\newblock {\tt arXiv:1112. 5840 [cs.OH]}, (2011).

\bibitem{Bergstra2012a}
J.A.\ Bergstra.
\newblock Decision taking as a service.
\newblock {\tt arXiv:1205.4194  [cs.SE]}, (2012).

\bibitem{Bergstra2012b}
J.A.\ Bergstra.
\newblock Decision taking versus action determination.
\newblock {\tt arXiv:  1205.6177v1  [cs.SE]}, (2012).

\bibitem{BDV2011a}
J.A.\ Bergstra, G.P.A.J.\ Delen, and S.F.M.\ van Vlijmen.
\newblock Outsourcing competence.
\newblock {\tt arXiv:1109.6536 [cs.SE]}, (2011).

\bibitem{BDV2011b}
J.A.\ Bergstra, G.P.A.J.\ Delen, and S.F.M.\ van Vlijmen.
\newblock Stratified outsourcing theory.
\newblock {\tt arXiv:1110.1957 [cs.SE]}, (2011).

\bibitem{BergstraLoots2002a}
J.A.\ Bergstra and M.E.\ Loots.
\newblock Program algebra for sequential code.
\newblock {\em Journal of Logic and Algebraic Programming}, 51 (2) pp. 125--156, (2002).

\bibitem{BergstraMiddelburg2007}
J.A.\ Bergstra and C.A.\ Middelburg.
\newblock Thread algebra for strategic interleaving.
\newblock {\em Formal Aspects of Computing}, 19(4) pp. 445--474, (2007).

\bibitem{BergstraMiddelburg2011}
J.A.\ Bergstra and C.A.\ Middelburg.
\newblock Thread algebra for poly-threading.
\newblock {\em Formal Aspects of Computing}, 23 pp. 567--583, (2011).

\bibitem{BergstraMiddelburg2012}
J.A.\ Bergstra and C.A.\ Middelburg.
\newblock Instruction sequence processing operators.
\newblock {\em Acta Informatica}, 49 pp. 139--172, (2012).

\bibitem{BergstraMiddelburg2012b}
J.A.\ Bergstra and C.A.\ Middelburg.
\newblock Instruction sequences for computer science.
\newblock {\em Atlantis Studies in Computing, Atlantis Press}, 
ISBN (print) 978-94-91216-64-0, (e-book) 978-94-91216-65-7, ISSN 221-8565 (2012).

\bibitem{BergstraPonse2007a}
J.A.\ Bergstra, and A.\ Ponse.
\newblock Execution architectures for program algebra.
\newblock {\em Journal of Applied Logic}, 5 (1) pp. 170--192, (2007).

\bibitem{BergstraPonse2011a}
J.A.\ Bergstra and A.\ Ponse.
\newblock Proposition algebra.
\newblock {\em ACM Transactions on Computational Logic}, Vol. 12 (3) Article 31 (36 pages), (2011).

\bibitem{BiyalogorskyGerstner2004}
E.\ Biyagolorsky and E.\ Gerstner.
\newblock Contingent pricing to reduce price risks.
\newblock {\em  Marketing Science}, Vol. 23 (1) pp. 146--155, (2004).

\bibitem{Burgess2005}
M. Burgess.
\newblock An approach to understanding policy based on autonomy and voluntary cooperation.
\newblock in:  {\em Ambient Networks,  Springer LNCS,} Vol 3775 pp. 97--108, (2005).

\bibitem{Burgess2007}
M. Burgess.
\newblock System administration and the scientific method.
\newblock in: J.A. Bergstra and M. Burgess (editors), {\em Handbook of Network and System Administration,}
\newblock pp. 689--728, (2007).


\bibitem{ButlerHighhouse2000}
A.\ Butler and S.\ Highhouse.
\newblock Deciding to sell: the effect of prior inaction and offer source.
\newblock {\em Journal of Economic Psychology}, Vol. 21 pp. 223--232, (2000).

\bibitem{Camerer1998}
C.\ Camerer.
\newblock Bounded rationality in individual decision making.
\newblock {\em Experimental Economics}, 1 pp. 163--183, (1998).

\bibitem{ChonkoDJR2003}
L.B.\ Chonko, A.J.\ Dubinsky, E.\ Jones, and J.A.\ Roberts.
\newblock Organizational and individual learning in the sales force: an agenda for sales research.
\newblock {\em Journal of Business Research}, Vol. 56  pp. 935--946, (2003).

\bibitem{CohenMO1972}
D.K.\ Cohen, J.G.\ March, and J.P.\ Olsen.
\newblock A garbage can model of organizational choice.
\newblock {\em Administrative Science Quarterly} Vol. 17  pp. 1--25, (1972).

\bibitem{Cribbet1986}
J.E.\ Cribbet.
\newblock Concepts in transition: the search for a new definition of property.
\newblock {\em University of Illinois Law Review}, Vol. 1986 (1)  42 pages, (1986).

\bibitem{GenesoveMayer2001}
D.\ Genesove and C.\ Mayer.
\newblock Loss aversion and seller behavior: evidence from the housing market.
\newblock {\em The Quarterly Journal of Economics}, Vol. 16 (4). pp. 1233--1260, (2001).

\bibitem{Goldberg1982}
D.S.\ Goldberg.
\newblock Fair market value in tax law: replacement value or liquidation value.
\newblock {\em Texas Law Review}, Vol. 60 (5) pp. 833--873, (1982).


\bibitem{HammondKR1998}
J.S.\ Hammond, R.J.\ Keeney, and H.\ Raiffa.
\newblock The hidden traps in decision making.
\newblock {\em Harvard Business Review}, reprint number 98505, (1998).

\bibitem{HaurinHNS2010}
D.R.\ Haurin, J.L.\ Haurin, T.\ Nadauld, and A.\ Sanders.
\newblock List prices, sales prices and marketing time: an application to U.S. housing markets.
\newblock {\em Real Estate Economics}, Vol. 38 (4) pp. 659--685, (2010).

\bibitem{Hodgkinson1997}
G.P.\ Hodgkinson.
\newblock Cognitive inertia in a turbulent market: the case of UK residential estate agents.
\newblock {\em Journal of Management Studies}, Vol. 34 (6) pp. 922--945, (1997).

\bibitem{Howard1988}
R.A.\ Howard.
\newblock Decision analysis: practice and promise.
\newblock {\em Management Science}, 34 (6) pp. 679--695, (1998).

\bibitem{HummelHNHKL2012}
H.G.K.\ Hummel, J.\ van Houcke, R.J.\ Nadolski, T.\ van der Hiele, H.\ Kurvers, and A.\ L\"{o}hr.
\newblock Scripted collaboration in serious gaming for complex learning: 
effects of multiple perspectives when acquiring water management skills.
\newblock {\em British Journal of Educational Technology}, Vol. 42 (6) pp. 1029--1041, (2012).

\bibitem{Johnson2006}
R.H.\ Johnson.
\newblock Making sense of ``Informal Logic''.
\newblock {\em Informal Logic}, Vol. 26 (3) pp. 231--258, (2006).

\bibitem{KimBKN2008}
A.K.\ Kim, D.\ Brooks, H.\ Kim, and J.\ Nissly.
\newblock Structured decision making and child welfare service delivery project.
\newblock {\em University of California at Berkeley, California Social Work Center},  (2008).

\bibitem{Knight2002}
J.R.\ Knight.
\newblock Listing price, time on market, and ultimate selling price: causes and effects of listing price changes.
\newblock {\em Real Estate Economics}, Vol. 30 (2) pp. 213--237, (2002).

\bibitem{Krainer2001}
J.\ Krainer.
\newblock A theory of liquidity in residential reals estate markets.
\newblock {\em Journal of Urban Economics}, Vol. 49 pp. 32--53, (2001 ).

\bibitem{LeighMarshall2001}
T.W.\ Leigh and G.W.\ Marshall.
\newblock Research priorities in sales strategy and performance.
\newblock {\em The Journal of Personal Selling and Sales Management}, Vol. 21 (2) pp. 83--93, (2001).

\bibitem{LeighMcGraw1989}
T.W.\ Leigh and P.F.\ McGraw.
\newblock Mapping the procedural knowledge of industrial sales personnel: 
a script-theoretic approach.
\newblock {\em Journal of Marketing}, Vol. 53 (1) pp. 16--34, (1989).

\bibitem{Middelburg2011}
C.A.\ Middelburg.
\newblock A survey of paraconsistent logics.
\newblock {\tt arXiv:1103.4324 [cs.LO]}, (2011).

\bibitem{Oakes1990}
G.\ Oakes.
\newblock The sales process and the paradoxes of trust.
\newblock {\em Journal of Business Ethics}, Vol. 9 pp. 671--679, (1990).

\bibitem{Rapp2009}
A.\ Rapp.
\newblock Outsourcing the sales process: hiring a mercenary sales force.
\newblock {\em Industrial Marketing Management}, Vol. 38 pp. 411--418, (2006).

\bibitem{SirmansTD1995}
C.F.\ Sirmans, G.K.\ Turnbull, and J.\ Dombrow.
\newblock Quick house sales: seller mistake or luck?
\newblock {\em Journal of Housing Economics}, Vol. 4 pp. 230--243, (1995).

\bibitem{SpiroWeitz1990}
R.L.\ Spiro and B.A.\ Weitz.
\newblock Adaptive selling: conceptualization, measurement, and nomological validity.
\newblock {\em Journal of Marketing Research}, Vol. 27 (1) pp. 61--69, (1990).

\bibitem{Valdman2010}
M.\ Valdman.
\newblock Outsourcing self-government.
\newblock {\em Ethics}, Vol. 120. (4) pp. 761--790 (2010).

\bibitem{ZeeHR2012}
D.-J.\ van der Zee, B.\ Holkenborg, and S.\ Robinson.
\newblock Conceptual modeling for simulation-based serious gaming.
\newblock {\em Decision Support Systems}, doi:10.1016/j/dss.2012.03.006 (2012).

\bibitem{ZeelenbergNPD2006}
M.\ Zeelenberg, B.A.\ Nijstad, M.\ van Putten, and E. van Dijk.
\newblock Inaction inertia, regret, and valuation: a closer look.
\newblock {\em Organizational Behavior and Human Decision Processes}, Vol. 101 pp. 89--104, (2006).


\end{thebibliography}
\end{document}